Imperial College London

Department of Computing

# OPESCI-FD: Automatic Code Generation Package for Finite Difference Models

*Author:*
Tianjiao Sun

*Supervisors:*
Dr Gerard Gorman
Prof Paul Kelly

Submitted in partial fulfilment of the requirements for the MSc degree in Computing of Imperial College London

May 20, 2016

This page intentionally left blank.

# Abstract

In this project, we introduce OPESCI-FD, a Python package built on symbolic mathematics to automatically generate Finite Difference models from a high-level description of the model equations. We investigate applying this framework to generate the propagator program used in seismic imaging. We implement the 3D velocity-stress FD scheme as an example and demonstrate the advantages of usability, flexibility and accuracy of the framework. The design of OPESCI-FD aims to allow rapid development, analysis and optimisation of Finite Difference programs. OPESCI-FD is the foundation for continuing development by the OPESCI project team, building on the research presented in this report. This report concludes by reviewing the further developments that are already under way, as well as the scope for extension to cater for other equations and numerical schemes.

# Acknowledgements

Foremost I would like to thank my supervisors, Dr Gerard Gorman and Prof Paul Kelly, for their expertise, enthusiasm and generous support throughout the project. I am extremely grateful for their confidence in trusting me with this project. This experience has opened the door to the exciting field of scientific computing for me.

OPESCI-FD is part of the OPESCI (Open Performance portablE SeismiC Imaging) project. The development of OPESCI-FD has benefited tremendously from the agile development approach of the overall project, with continuous integration established from day one. Throughout the project, the author has continued to benefited from discussions with other members of the OPESCI team. I would like to thank:

Dr Michael Lange, for building the benchmark suit of OPESCI-FD, and for his help in reorganising the repository into a Python package (which is not nearly as easy as it sounds).

Dr Christian Jacobs, for reviewing my code, and for many interesting discussions about numerical methods in geophysics.

Marcos and Felippe from SENAI CIMATEC in Salvador, Brazil, for their help in testing and collecting results from the Xeon Phi platform.

Thank you team, it has been a pleasure working with you guys!

Lastly but most importantly, I would like to thank my wonderful wife, Yu, for her love, patience and support. It was a difficult choice for me to go back to college last summer and pursue what I want for my life. Her unreserved trust and support allowed me to enjoy my study so much, and I am truly fortunate to have found her.

# Contents

# Chapter 1

# Introduction

## 1.1 Motivation

Partial differential equations (PDEs) form the basis of a large range of mathematical models describing physical, chemical and biological phenomena, and more recently their use has spread into economics, financial forecasting, image processing and many other fields [44][59]. Despite being ubiquitous, only a small subset of PDEs have analytical solutions and usually only for special cases. To solve real systems of scientific and engineering interests, it is usually necessary to solve the PDEs numerically. The phenomenal advance in high-performance computing (HPC) enables more sophisticated numerical methodologies to be applied. However, implementation of these algorithms manually is often an laborious and error-prone process. Code reusability is also hindered due to the lack of common abstraction among different algorithms. Development of new schemes often starts from scratch.

Figure 1.1 shows a typical work flow for applying PDEs in scientific computing and the different research disciplines involved. It starts with natural scientists establishing the governing equations for the system of interests. Numerical analysts derive numerical discretisation (often referred to as schemes or algorithms) to solve the PDEs. Finally, this is implemented in software to run on computers. Throughout the decades several well-studied numerical methods have emerged and been applied successfully to a wide range of problems. These include the Finite Difference Method (FDM), Finite Element Method (FEM) and Finite Volume Method (FVM).

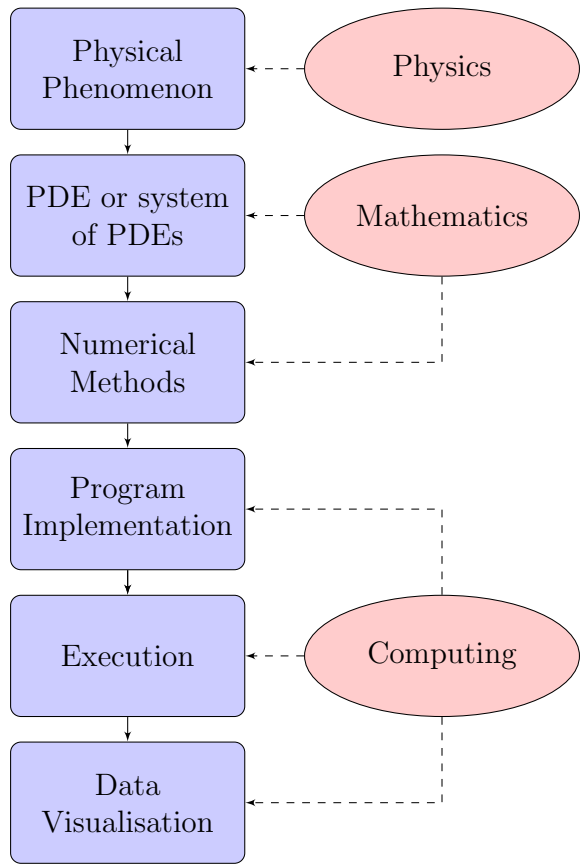

Figure 1.1: Typical workflow of numerical solution of PDE

The opportunity of computer science contribution is in three stages of this process: 1) to implement the numerical schemes in the programming languages of choice, 2) to optimise the performance of such code under specific hardware and software constrains, and 3) to process the input and output data in the most useful format.

In this project, we developed OPESCI-FD (Open Performance portablE SeismiC Imaging - Finite Difference), a Python software package that automates the transformation of mathematical models into source code using finite differences discretisation. It provides higher level symbolic mathematics abstraction so that the application developers (such as geophysicists in this case study) are alleviated from implementation details. This automation aims to improve productivity, reduce programming errors, and enable systematic fine-tuning of the solvers to achieve high performance on modern many-core architectures. We limit the initial scope of our study to FD schemes for the elastic wave equation used in geological exploration as our motivating example.

This project is divided into two major components, graphically represented in Figure 1.2. The first phase focuses on discretisation of the PDEs and create the computational kernel (the computation to update the domain at each time step) using the FD approximations. OPESCI-FD builds on the SymPy[1] library of Python programming language to create symbolic mathematics routines to achieve this transformation automatically. For 3D problems using high order discretisations, the mathematical kernels can be large and complex. This issue clearly illustrates the benefit of automating this process, as even the task of typing out these very long expressions manually is highly time-consuming and error-prone. In the next phase OPESCI-FD transforms the FD representation into compilable source code. We exploit the structural regularity of the finite difference schemes, which allows using templates as blueprints of the target source code. We start by generating sequential C/C++ code, with the aim to extend to OpenMP API and other stencil compilers and domain-specific languages (DSLs).

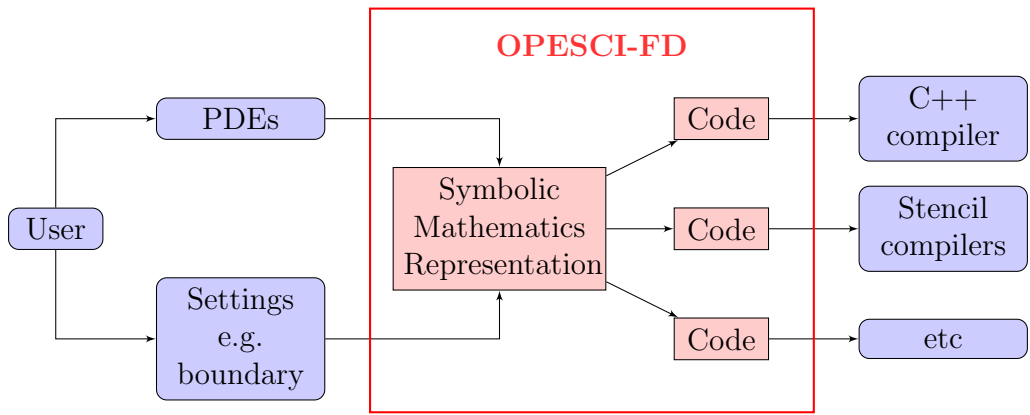

Figure 1.2: Schematic representation of OPESCI-FD work flow. OPESCI-FD transforms high level description of the problem into symbolic mathematics representation, which is then converted to source code for different downstream compilers.

---

[1] http://www.sympy.org

## 1.2 Objectives

The core objective of OPESCI-FD is to allow rapid development of FD equations used in seismic imaging. Whilst there are multiple FD schemes used in the field of seismic imaging, OPESCI-FD aims to implement the most representative approaches of the problem, and provide flexibility for the application developer to adapt the scheme easily. In doing so, we attempt to present the framework of using symbolic mathematics to facilitate scientific computing as a general concept.

## 1.3 Contribution

The contribution of this project is two-fold:

- On the macro level, with the techniques shown in this project, we propose a framework of solving PDEs using code generation aided by symbolic mathematics. OPESCI-FD demonstrates many advantages of such approach over manual implementation of FD schemes, such as usability, flexibility and accuracy. We also explore the possibility of code generation targeting different back-end stencil tools in the same framework.

- On the micro level, OPESCI-FD successfully implemented one of the most widely used FD schemes in geophysics, namely the velocity-stress scheme on a staggered grid. This application demonstrates the potential of the framework. We recognise that OPESCI-FD will benefit greatly from further developments to handle more FD schemes in other areas of scientific computing, as discussed in Section 7.2.

OPESCI-FD is an open source-project[2], designed to be highly modularised and extensible to encourage contributions. The project is under active development with new features added frequently. This report corresponds to commit number 131 of the repository.

---

[2]https://github.com/opesci/opesci-fd

## 1.4 Report outline

The rest of this report is organised as follows:

Chapter 2 looks into the existing techniques used in computational seismic studies, with emphasis on FD methods. We also briefly survey the different approaches of applying symbolic mathematics to scientific computing, as well as the development in stencil tools which could be used to solve FD schemes.

Chapter 3 illustrates, with the simple example of 1D wave equation, the basic techniques that OPESCI-FD applied in derivation of the mathematical kernels and code generation.

Chapter 4 discusses some of the most frequently used FD schemes in geophysics, explores the common themes that OPESCI-FD can exploit to build a higher level of abstraction. Again examples in 1D are used, but the ideas extend to higher dimensions.

Chapter 5 shows the key features and components of the design of OPESCI-FD. We discuss the design choices we made in order to achieve a flexible and extensible framework.

Chapter 6 presents implementation details of OPESCI-FD by applying OPESCI-FD to build a FD solver that resembles real world applications. This example demonstrates the work flow of OPESCI-FD and the advantages of the framework.

Finally, Chapter 7 provides a summary of this project, the current state of OPESCI-FD development, and points to possible direction of future works.

This page intentionally left blank.

# Chapter 2

# Background

In this chapter we discuss the background and related work of this project. Section 2.1 provides a brief historical review of the development of FDM applied in seismic modelling. Section 2.2 summarises the techniques of using symbolic mathematics to generate code in scientific computing. Section 2.3 surveys the stencil tools developed for scientific computing, with the aim of selecting suitable back-end compilers for OPESCI-FD.

## 2.1 Finite Difference Method in seismic modelling

In this section, we give a brief historical account of applying FDM in seismic modelling. In particular, we focus on the development of velocity-stress schemes on staggered grid, which is the dominating scheme used in industry and has been implemented in OPESCI-FD. More comprehension reviews of this topic can be found in [42] and [24].

The early application of FDM to seismic wave propagation was based on the displacement formulation on regular grid[1]. Early pioneering studies include [1], [9], [29] among many others. Later researches, such as [33], [60], [43] attempted to improve the accuracy and performance of this formulation until the end of 1990s. These studies achieved reasonable accuracy in modelling planner free-surface boundary conditions, but could not overcome the fundamental limitation of the displacement FD formulation: The displacement schemes are prone to introduce instability and numerical dispersion under the situation of large P-wave to S-wave speed ratio ($V_p/V_s$, known as Poisson's ratio) [42].

---

[1]More details can be found in Chapter 4.

This problem with the displacement formulation led Madariaga [40] [41] to introduce velocity-stress scheme on staggered grid in 1970s, initially to model frictional faults. This was motivated by Yee's work on electromagnetics in his seminal 1966 paper [57]. The novelty of this formulation is to apply centred FD approximations in both space and time on staggered grid for each vector field (electric and magnetic field in Yee's case, and velocity and stress field in Madariaga's case). Virieux introduced the 2D velocity-stress scheme of second-order accuracy in both space and time for modelling SH waves [54], and later P-SV waves [55]. In these works, Virieux also showed that stability condition of P-SV wave models is independent of Poisson's ratio in this scheme. This is a big advantage over regular grid based schemes, and soon staggered grid based schemes become the dominant type of FD formulation in seismic studies till this day. To enhance the computational efficiency, later works strived to improve the approximation accuracy in order to reduce the the number of grid points required. Levander [35] introduced the fourth-order spatial approximation to the scheme. Yomogida and Etgen [58] developed displacement-stress scheme with eighth-order approximation in space on staggered grid. Graves [22] extended the scheme to 3D.

Other schemes and grids have been introduced over the decades. Examples include, but not limited to, displacement-stress scheme on spatially staggered grid [39], velocity-stress scheme on partly staggered grid [4], optimally accurate scheme [20]. More recently, the standard staggered grid has been generalised to Lebedev grid [60], which is possibly a natural choice for an anisotropic medium. Collocated grid [61] has also been introduced for velocity-stress schemes to overcome the difficulty in imposing boundary conditions on staggered grid.

## 2.2 Application of symbolic mathematics in code generation for scientific computing

Problems in scientific computing, such as solving PDEs, normally have compact mathematical representations. Techniques in symbolic mathematics have been applied in this domain to exploit the intrinsic mathematical properties of the problems. Symbolic computation specialise in the exact computation of numbers, symbols, vectors and the like, and is therefore complementary to numerical methods using floating point numbers.

Early works by Cecchi and Lami [11], Korncoff and Fenves [31], and Noor and Anderson [45] demonstrated the potential of applying symbolic computation to scientific computing. The scope then was limited to generating some parts of the implementation, such as the coefficients of the FEM formulae, rather than generating fully compilable source code. Wang [56] developed FINGER, a LISP based software to derive formulae needed in finite element analysis. The key step in FINGER is to derive the material property matrix with symbolic matrix computation. The formulae can be translated to FORTRAN code. Korelc [30] developed a MATHEMATICA package to automatically derive the formulae needed in FEM. The formulae can then be output in C or FORTRAN format. Amberg *et al.* [2] created a toolbox embedded in Maple to solve PDEs with FEM, also by generating FORTRAN code, using the CodeGeneration package in Maple. More recently, the FEniCS project [36] provides a more complete tool-chain for automated solution of differential equations in Python. In particular, FEniCS introduces the Unified Form Language (UFL) [38] to allow users to express the PDEs close to their mathematical forms, and the FEniCS Form Compiler (FFC) [37] to automate code generation.

There is relatively little work on applying symbolic mathematics to generate FD PDE solvers.

## 2.3 Stencil compilers

Several stencil languages have been proposed to perform computation on structured grids, where the value at each grid point is updated based on the values of its neighbouring points. The construction of these systems vary from "black box" environments to C++ libraries. The most important task of such tools is to exploit parallelism of problems to generate source code which is optimised for the target hardware platform. Pochoir [52] compiler employs divide-and-conquer technique to generate cache-oblivious Cilk code. It provides clean abstraction to describe the kernel and boundary conditions based on C++ Standard Template Library (STL), but the rigid structure restricts the choices of schemes when applied to vector field problems. Lizst [18] is an external DSL to construct mesh-based solvers. It exposes the parallelism by analysing language statements, and generates code to choice of platform, such as MPI, OpenMP, and CUDA. PATUS [12] generates C source code for stencil computation on shared-memory CPU architectures. The compiled executable can be passed to its auto-tuning tool to determine a set of architecture-specific parameters for best performance. All these stencil languages offer simple mechanism to address neighbouring points. This eases expressing dependency in regular grid in FD applications. Polly [23] is one of the tools implementing polyhedral tiling algorithms, built on LLVM infrastructure. It is designed to work on LLVM-IR (intermediate representation) to LLVM-IR transformation, so theoretically no change in the C++ source code is required to apply Polly optimisations.

Some projects mainly target applications on unstructured mesh but can be applied on structured grid as well. OP2 [21] takes the Æcute (access-execute descriptor) programming approach that separates the specification of an element-wise computational kernel with its parallel invocations. This design gives the library freedom to choose how to parallelise the operation. PyOP2 [50] is based on OP2 and uses Python as the host language. It targets mesh-based simulation codes over unstructured meshes to achieve parallel execution of computational kernels on multicore CPUs and GPUs. Furthermore, it also employs JIT (just-in-time) compilation and parallel scheduling.

Other approaches focus on providing a framework consisting of front end and back end modules to support stencil application development. Project ExaStencils [51] provides an infrastructure for programming numerical solvers in a DSL called ExaSlang, targeting distributed memory systems. It uses ployhedral model for loop optimisation and generates C++ code utilizing a hybrid OpenMP and MPI parallelisation. ROSE [47] provides compiler infrastructure for building source-to-source analysis or optimization tools for C/C++/Fortan codes by providing the corresponding front ends, back ends, and data structures. Its loop optimiser implements multi-level fusion algorithms to perform dependency analysis. DUNE [6] is a software framework for the numerical solution of PDEs with grid-based methods. It supports reuse of existing numerical packages such as UG [5], Alberta [10], and ALU3d [16]. Overture [26] is an object-oriented code framework for developing PDE solvers. It provides abstraction for moving components with complex geometry.

Mint [53] and STELLA [46] are DSLs targeting stencil codes on structured grids. STELLA is embedded in C++ through template meta-programming, it provides two back ends, one based on OpenMP for multi-core CPUs, and one based on CUDA for GPUs. Data exchange for distributed memory architecture is supported by Generic Communication Library (GCL [7]). PLuTo [8] is an automatic source-source (sequential C to OpenMP) transformation tool based on polyhedral model, which uses linear optimisation with respect to parallelism and data locality to search for tiling transformation choices.

Similar to the approach of OPESCI-FD, some frameworks focus on expressing the PDEs at the mathematical level which allows intuitive symbolic manipulation before applying numerical methods. Sandia's Sundance [38], FreeFEM++ [25], Firedrake [49] and FEniCS [19] use this higher level of abstraction that expresses the problems in terms of actual differential equations, leaving the details of the parallelisation to a lower-level library. In particular, Firedrake and FEniCS use Unified Form Language (UFL) to express variational forms for finite element analysis. OPESCI-FD shares this high level automation philosophy, with the aim of targeting a selection of downstream stencil tools.

## 2.4  Related work

OPESCI-FD builds on previous research such as [56], [31], [2] on using symbolic mathematics and code generation techniques to solve PDEs. The work is also inspired by the use of UFL in tools like Firedrake. OPESCI-FD is an object-oriented framework created around existing Python libraries, and makes heavy use of template files to create compilable source code. Unlike the above mentioned work, OPESCI-FD is targeting FDM rather than FEM algorithms. One key design principle of OPESCI-FD is to maintain flexibility to generate code to different downstream stencil tools. The optimisation strategy in OPESCI-FD is therefore focused on helping the downstream compilers with flags, directives and structural features in the generated source code to produce high-performing implementation.

In this chapter we have discussed the context and needs of OPESCI-FD. Next we will demonstrate the basic techniques used by OPESCI-FD to generate the computational kernels using a simple 1D example.

# Chapter 3

# Case study: wave equation in 1D

In this chapter we use the one-dimensional wave equation to illustrate the techniques used in OPESCI-FD. In particular we discuss the following key steps:

- Derivation of computational kernel using symbolic mathematics.

- Code fragment generation from symbolic mathematical representation.

- Insertion of code fragments into templates to generate compilable source code.

These steps form the building blocks when we solve more complicated problems in following chapters. The overall strategy stays the same for different problems and schemes. It is also worth noting that each of these steps can be extended independently, and assembled together later. This process is more profound and powerful than simply linking libraries due to the interactions among different components. The overall work flow is therefore highly modularised and flexible.

## 3.1 Finite Difference formulation

Wave equations are associated with many physical phenomena and are among the most thoroughly studied PDEs. The wave equation in 1D is defined as

$$\frac{\partial^2 u}{\partial t^2} = c^2 \frac{\partial^2 u}{\partial x^2}, \qquad x \in [0, L], t \in [0, T], \tag{3.1}$$

where $u$ is the displacement, $x$ is coordinate in 1D, $t$ is time, $L$ and $T$ define the space and time domain, and constant $c$ is the speed of wave propagation in the medium. The wave equation in 1D was inspired by studying small vibrations on a string under tension. It was first derived and solved analytically by Jean le Rond d'Alembert in 1747 [15]. Here we demonstrate the process of solving this equation with finite difference method, in order to identify the key steps to be automated in OPESCI-FD. We will use a more compact notation for the partial derivatives:

$$u_t = \frac{\partial u}{\partial t}, \qquad u_{tt} = \frac{\partial^2 u}{\partial t^2} \tag{3.2}$$

and similar expressions for derivatives with respect to other variables. Then the wave equation can be written compactly as $u_{tt} = c^2 u_{xx}$.

### 3.1.1 Discretisation of the domain

The temporal domain $[0, T]$ is discretised into a finite number of *time levels*:

$$0 = t_0 < t_1 < t_2 < \cdots < t_{N_t-1} < t_{N_t} = T \tag{3.3}$$

Similarly, the spatial domain $[0, L]$ is discretised into a set of grid points:

$$0 = x_0 < x_1 < x_2 < \cdots < x_{N_x-1} < x_{N_x} = T \tag{3.4}$$

For uniformly distributed mesh points we can introduce the constant mesh spacings $\Delta x$ and $\Delta t$. We thus have:

$$\begin{aligned} x_i &= i\Delta x, & i &= 0, \ldots, N_x \\ t_i &= n\Delta t, & n &= 0, \ldots, N_t \end{aligned} \tag{3.5}$$

It is helpful to view the mesh points as a regular two-dimensional grid in the $x, t$ plane, consisting of points $(x_i, t_n)$, with $i = 0, \ldots, N_x$ and $n = 0, \ldots, N_t$, as shown in Figure 3.1.

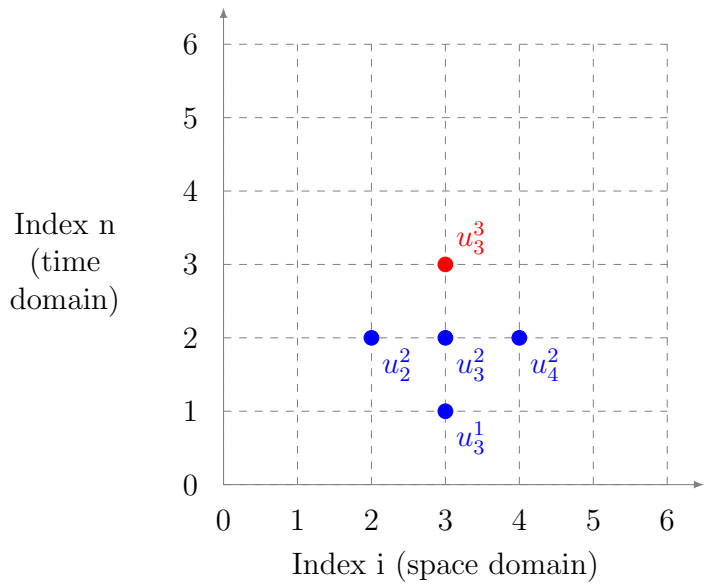

Figure 3.1: Finite difference grid for a 1D wave equation.

### 3.1.2 The discrete solution at grid points

We introduce the notation $u_i^n$ as the numerical value at grid point $(x_i, t_n)$ to approximate the exact solution $u(x,t)$ in Equation (3.1). For a numerical solution by the finite difference method, we relax the condition that (3.1) holds at all points in the space-time domain $[0, L] \times [0, T]$ to the requirement that the PDE is fulfilled at the interior grid points:

$$\frac{\partial^2}{\partial t^2} u(x_i, t_n) = c^2 \frac{\partial^2}{\partial x^2} u(x_i, t_n), \qquad (3.6)$$

for $i = 1, \ldots, N_x - 1$ and $n = 1, \ldots, N_t - 1$. For the points at the borders ($n = 0, x = 0, x = N_x$) we need to incorporate the initial conditions and boundary conditions of the specific problems, as illustrated in Section 3.1.6.

### 3.1.3 Replacing derivatives with finite differences

We start by considering the space domain. The first order derivative in $x$ can be approximated by *difference quotient* for small $\Delta x$ as

$$\frac{\partial}{\partial x} u(x,t) \approx \frac{u(x+\Delta x, t) - u(x,t)}{\Delta x}. \tag{3.7}$$

Approximations such as equation (3.7) are known as the *forward difference* form of finite differences. Depending on the application, one can also choose the *central difference* form:

$$\frac{\partial}{\partial x} u(x,t) \approx \frac{u(x+\Delta x, t) - u(x-\Delta x, t)}{2\Delta x}, \tag{3.8}$$

or written more compactly, at mesh point $(x_i, t_n)$, we have

$$u_x(x_i, t_n) \approx \frac{u_{i+1}^n - u_{i-1}^n}{2\Delta x}. \tag{3.9}$$

Similarly, the second order derivatives can be approximated as the difference quotient of different quotients:

$$\begin{aligned}
\frac{\partial^2}{\partial x^2} u(x,t) &\approx \frac{1}{\Delta x}\left[\left.\frac{\partial u}{\partial x}\right|_{x+\Delta x} - \left.\frac{\partial u}{\partial x}\right|_x\right] \\
&\approx \frac{1}{\Delta x}\left[\frac{u(x+\Delta x, t) - u(x,t)}{\Delta x} - \frac{u(x,t) - u(x-\Delta x, t)}{\Delta x}\right] \\
&= \frac{1}{(\Delta x)^2}\left[u(x+\Delta x, t) + u(x-\Delta x, t) - 2u(x,t)\right].
\end{aligned} \tag{3.10}$$

Thus at mesh point $(x_i, t_n)$, we have

$$u_{xx}(x,t) \approx \frac{u_{i+1}^t + u_{i-1}^t - 2u_i^t}{\Delta x^2}. \tag{3.11}$$

The goal of FD formulation is to express derivatives with functions of values at neighbouring grid points, such as Equations (3.9) and (3.11). A systematic approach to derive such expressions is through Taylor series expansion. We first write down the Taylor series expansions at point $(x_{i-1}, t_n)$ and $(x_{i+1}, t_n)$ with respect to $(x_i, t_n)$. For the ease of reading, in this section we omit the coordinates of the partial derivatives, so that we use $u_{xx}$ as a shorthand for $u_{xx}(x_i, t_n)$ and similarly for derivatives of other orders.

$$u_{i-1}^n \approx u_i^n - \frac{\Delta x}{1!} u_x + \frac{(\Delta x)^2}{2!} u_{xx} - \frac{(\Delta x)^3}{3!} u_{xxx} + O(\Delta x^4), \qquad (3.12)$$

$$u_{i+1}^n \approx u_i^n + \frac{\Delta x}{1!} u_x + \frac{(\Delta x)^2}{2!} u_{xx} + \frac{(\Delta x)^3}{3!} u_{xxx} + O(\Delta x^4). \qquad (3.13)$$

Take the difference of (3.12) and (3.13), rearrange and divide through by $\Delta x$, we obtain

$$u_x \approx \frac{u_{i+1}^n - u_{i-1}^n}{2\Delta x} + O(\Delta x^2). \qquad (3.14)$$

Take the sum of the same equations (3.12) and (3.13), rearrange and divide through by $\Delta x^2$, we obtain

$$u_{xx} \approx \frac{u_{i+1}^t + u_{i-1}^t - 2u_i^t}{\Delta x^2} + O(\Delta x^2). \qquad (3.15)$$

Truncating higher order terms off Equations (3.14) and (3.15), we arrive at the previous results (3.9) and (3.11).

More conveniently, we can use matrix algebra to represent the above relationship between the partial derivatives at one specific grid point and the Taylor series terms at its neighbouring points as

$$\underbrace{\begin{bmatrix} u_{i-1}^t \\ u_i^t \\ u_{i+1}^t \end{bmatrix}}_{U} = \underbrace{\begin{bmatrix} 1 & -\frac{\Delta x}{1!} & \frac{\Delta x^2}{2!} \\ 1 & 0 & 0 \\ 1 & \frac{\Delta x}{1!} & \frac{\Delta x^2}{2!} \end{bmatrix}}_{M} \times \begin{bmatrix} u_i^t \\ u_x \\ u_{xx} \end{bmatrix}. \qquad (3.16)$$

The derivatives can then be found by inverting the matrix $M$ and post-multiply by the vector $U$. Such algorithm can be implemented in a symbolic mathematics library such as Mathematica or SymPy. Below is the Python code fragment that achieve this using SymPy. Algorithm in Figure 3.3 creates the matrix $M$, algorithm in Figure 3.4 derives the FD approximations of derivatives.

Figure 3.2 shows the generated 2, 4, 12 order FD approximation of $\frac{\partial u}{\partial x}$. It is obvious that higher order approximations are increasingly complicated for human implementation. The algorithm can also be extended to other FD schemes, such as the forward difference form in (3.7). To do that we simply need to create matrix $M$ with different combination of neighbouring grid points.

```
>>> Deriv(U, [t,x], 1, dx, 2)[1]    # 2nd order
'-U[t,x-1]/(2*dx)+U[t,x+1]/(2*dx)'
>>> Deriv(U, [t,x], 1, dx, 4)[1]    # 4th order
'-U[t,x-3]/(60*dx)+3*U[t,x-2]/(20*dx)-3*U[t,x-1]/(4*dx)+3*U[t,x+1]/(4*dx)-3*U[t,x+2]/(20*dx)
    +U[t,x+3]/(60*dx)'
>>> Deriv(U, [t,x], 1, dx, 12)[1]   # 12th order
'-U[t,x-11]/(7759752*dx)+11*U[t,x-10]/(3527160*dx)-11*U[t,x-9]/(302328*dx)+55*U[t,x
    -8]/(201552*dx)-55*U[t,x-7]/(37128*dx)+11*U[t,x-6]/(1768*dx)-11*U[t,x-5]/(520*dx)+11*U[t
    ,x-4]/(182*dx)-55*U[t,x-3]/(364*dx)+55*U[t,x-2]/(156*dx)-11*U[t,x-1]/(12*dx)+11*U[t,x
    +1]/(12*dx)-55*U[t,x+2]/(156*dx)+55*U[t,x+3]/(364*dx)-11*U[t,x+4]/(182*dx)+11*U[t,x
    +5]/(520*dx)-11*U[t,x+6]/(1768*dx)+55*U[t,x+7]/(37128*dx)-55*U[t,x+8]/(201552*dx)+11*U[t
    ,x+9]/(302328*dx)-11*U[t,x+10]/(3527160*dx)+U[t,x+11]/(7759752*dx)'
```

Figure 3.2: Generated FD approximations of $\frac{\partial u}{\partial x}$ of different accuracy order

```python
from sympy import factorial, Matrix
def tc(dx, n):
    """
    return coefficient of power n term in Taylor series expansion
    :param n: power
    :param dx: distance from expansion reference
    """
    return (dx**n)/factorial(n)
def Taylor(dx, n):
    """
    create Matrix of Taylor Coefficients M
    such that M * D = R
    D is list of derivatives at x [f, f', f'' ..]
    R is list of values at neighbour grid point [..f(x-dx),f(x),f(x+dx)..]
    :param dx: spacing between grid points
    :param n: length of D and R (i.e. # of derivatives)
    returns Maxtrix object M
    """
    l = []
    for i in range(-n+1, n):
        ll = [tc((i*dx), j) for j in range(2*n-1)]
        l.append(ll)
    return Matrix(l)
```

Figure 3.3: Algorithm to create Taylor expansion matrix

```python
def Deriv(U, i, k, d, n):
    """
    calculate the FD approximation for nth derivative
    the function works by inverting M in M * D = R as defined in Taylor()
    :param U: the field
    :param i: list of indices (Symbol objects) of field U
    e.g. possibley [t,x,y,z] for 3D
    :param k: determine dependent variable
    e.g. k=0 for dU/dt, k=1 for dU/dx, assuming i=[t,x,y,z]
    :param d: spacing of the grid, e.g. dx
    :param n: order of accuracy of approximation
    e.g. n=2 for 2nd-order FD approximation
    returns D = list of expressions for FD approxiation
    """
    M = Taylor(d, n)
    s = [0]*len(i)
    s[k] = 1  # switch on kth dimension
    # generate matrix of RHS, [...U[x-1],U[x],U[x+1]...]
    if len(i) == 1:
        RX = Matrix([U[i[0]+s[0]*x] for x in range(-n+1, n)])
    elif len(i) == 2:
        RX = Matrix([U[i[0]+s[0]*x,i[1]+s[1]*x] for x in range(-n+1,n)])
    elif len(i) == 3:
        RX = Matrix([U[i[0]+s[0]*x,i[1]+s[1]*x,i[2]+s[2]*x]
                    for x in range(-n+1, n)])
    elif len(i) == 4:
        RX = Matrix([U[i[0]+s[0]*x,i[1]+s[1]*x,i[2]+s[2]*x,i[3]+s[3]*x]
                    for x in range(-n+1,n)])
    else:
        raise NotImplementedError(">4 dimensions, not implemented.")
    return M.inv() * RX
```

Figure 3.4: Algorithm to derive FD approximation of derivatives

### 3.1.4 Finite difference equation

A similar approximation of the second-order derivative in the time domain reads

$$u_{tt}(x,t) \approx \frac{u_i^{t+1} + u_i^{t-1} - 2u_i^t}{\Delta t^2}. \tag{3.17}$$

We can now replace derivatives in (3.1) with (3.15) and (3.17) to obtain the *difference equation*, also referred to as *algebraic equation* or *finite difference scheme*

$$\frac{u_i^{t+1} + u_i^{t-1} - 2u_i^t}{\Delta t^2} = c^2 \frac{u_{i+1}^t + u_{i-1}^t - 2u_i^t}{\Delta x^2}. \tag{3.18}$$

Rearrange to solve for $u_i^{t+1}$

$$u_i^{t+1} = -u_i^{t-1} + 2u_i^t + C^2(u_{i+1}^t - 2u_i^t + u_{i-1}^t), \tag{3.19}$$

where we introduce the dimensionless parameter

$$C = c\frac{\Delta t}{\Delta x}, \qquad (3.20)$$

known as the *Courant-Friedrichs-Lewy number*, or *CFL number*. It is the key parameter that governs the stability of the numerical solution [13]. Discussion of conditions on $C$ to ensure convergence of the FD scheme is beyond the scope of this project. However, we note that both the primary physical parameter of the medium, $c$, and the choice of numerical parameters $\Delta x$ and $\Delta t$ define $C$, which can be determined at runtime. One strength of our proposed method is to enable code generation for various combinations of numerical parameters such as $\Delta x$ and $\Delta t$, which improve efficiency while ensuring accuracy for a particular set of physical parameters of the problem. This is the starting point for automatic performance fine-tuning.

The term *stencil* is often used to describe the geometry aspect of the difference equation. The geometry of a typical stencil is illustrated as the coloured points in Figure 3.1. Here the function value at coordinate $(3,3)$, $u_3^3$, can be calculated from $u_2^2$, $u_3^2$, $u_4^2$, and $u_3^1$, according to equation (3.19).

### 3.1.5 Updating the mesh grid

At time $t = t_{n+1}$, the only unknown in Equation (3.19) is $u_i^{i+1}$. We can thus iterate through the grid in time domain, and for each time step, we then move the stencil across the space domain to update the value at each interior grid point with Equation (3.19). The below code fragment shows a typical nested loops implementation in C.

```c
for(int n=1; n<Nt; n++)
    for(int i=1; i<Nx; i++)
        // update mesh points at time = n+1
        u[i][n+1] = 2*u[i][n] - u[i][n-1]
                  - pow(C,2.0)*(u[i+1][n] - 2*u[i][n] + u[i-1][n])
```

By extending the CodePrinter facility in SymPy (see Chapter 5 for more details), we can convert the symbolic expression generated with algorithms described in Section 3.1.3 to the following C code. This code fragment can later be insert into a template file to form the body of the desired nested loops.

```
U[x][t + 1]=(-2*pow(dt, 2)*U[x][t] + pow(dt, 2)*U[x - 1][t] + pow(dt, 2)*U[x + 1][t] + pow(h
    , 2)*(2*U[x][t] - U[x][t - 1]))/pow(h, 2)
```

To maintain notation consistency when extending to multiple dimensions, in the generated source code, we substitute $\Delta x$ with $h$, $\Delta t$ with $dt$, $i$ with $x$, and $n$ with $t$ in the rest of the report.

### 3.1.6 Initial conditions and boundary conditions

Equation (3.1) contains second-order derivative in time domain, we therefore need two *initial conditions*. Below is one possible set of such conditions. Here (3.21) defines the initial shape of the string as function $I(x)$. (3.22) states that the initial velocity of the string is zero everywhere. Furthermore, equations (3.23) and (3.24) are *boundary conditions* specifying that the string is fixed at both ends.

$$u(x, 0) = I(x), \qquad x \in [0, L], \qquad (3.21)$$

$$\frac{\partial}{\partial t} u(x, 0) = 0, \qquad x \in [0, L], \qquad (3.22)$$

$$u(0, t) = 0, \qquad t \in [0, T], \qquad (3.23)$$

$$u(L, t) = 0, \qquad t \in [0, T], \qquad (3.24)$$

It is straightforward to modify the implementation we have described with conditions (3.21),(3.23) and (3.24). To account for the initial velocity condition (3.22), we observe from the first-order FD approximation in time that

$$u_t \approx \frac{u_i^{n+1} - u_i^{n-1}}{2\Delta t} = 0, \qquad \text{when } n = 0 \qquad (3.25)$$

$$\text{Therefore,} \qquad u_i^{n+1} = u_i^{n-1}, \qquad \text{when } n = 0 \qquad (3.26)$$

Substituting this derived condition (3.26) into the difference equation (3.19), we obtain a new expression to calculate the value for the first time step

$$u_i^1 = u_i^0 + \frac{1}{2}C^2(u_{i+1}^0 + u_{i-1}^0 - 2u_i^0). \tag{3.27}$$

Below is the modified implementation in C

```c
// initialisation
for(int i=0; i<=Nx; i++)
    u[i][0] = I(x[i])
// separate calculation for first time step
for(int i=1; i<Nx; i++)
    u[i][1] = u[i][0] - 0.5*pow(C,2.0)*(u[i+1][0] - 2*u[i][0] + u[i-1][0])
// main loop
for(int n=1; n<Nt; n++){
    // boundary conditions
    u[n+1][0] = 0; u[n+1][Nx] = 0;
    for(int i=1; i<Nx; i++)
        // update grid points at time = n+1
        u[i][n+1] = 2*u[i][n] - u[i][n-1] - pow(C,2.0)*(u[i+1][n]
                    - 2*u[i][n] + u[i-1][n])
}
```

## 3.2 Wave equations in 2D and 3D

In $n$-dimensional space, with spatial variables $x_1, x_2, \ldots, x_n$, and time variable $t$, the wave equation for scalar function $u = u(x_1, x_2, \ldots, x_n)$ is:

$$\frac{\partial^2 u}{\partial^2 t} = c^2 \nabla^2 u, \tag{3.28}$$

where $\nabla^2$ is the spatial Laplace operator, and $c$ is a scalar constant. Using $(x, y, z)$ as spatial variables in the Cartesian space, we can write the wave equation in two-dimensional and three-dimensional space as

$$\frac{\partial^2 u}{\partial^2 t} = c^2 \left( \frac{\partial^2 u}{\partial x^2} + \frac{\partial^2 u}{\partial y^2} \right), \tag{3.29}$$

$$\frac{\partial^2 u}{\partial^2 t} = c^2 \left( \frac{\partial^2 u}{\partial x^2} + \frac{\partial^2 u}{\partial y^2} + \frac{\partial^2 u}{\partial z^2} \right). \tag{3.30}$$

We then follow the same steps to replace partial derivatives as FD approximations. Here we choose $\Delta x = \Delta y = \Delta z = h$, and apply the previous algorithms to generate the following kernel for 2D wave equation solver in C. This implementation is second order accurate in both space and time, because the leading term of truncation error is $\Delta x^2$ in Equation (3.15) and $\Delta t^2$ in Equation (3.17).

```
U[x][y][t + 1] = (-4*pow(dt, 2)*U[x][y][t] + pow(dt, 2)*U[x][y - 1][t]
               + pow(dt, 2)*U[x][y + 1][t] + pow(dt, 2)*U[x - 1][y][t]
               + pow(dt, 2)*U[x + 1][y][t] + pow(h, 2)*(2*U[x][y][t]
               - U[x][y][t - 1]))/pow(h, 2)
```

And similarly for the 3D case, the kernel code is

```
U[x][y][z][t+1]=(-6*pow(dt, 2)*U[x][y][z][t] + pow(dt, 2)*U[x][y][z-1][t]
        + pow(dt, 2)*U[x][y][z + 1][t] + pow(dt, 2)*U[x][y - 1][z][t]
        + pow(dt, 2)*U[x][y + 1][z][t] + pow(dt, 2)*U[x - 1][y][z][t]
        + pow(dt, 2)*U[x + 1][y][z][t]
        + pow(h, 2)*(2*U[x][y][z][t] - U[x][y][z][t - 1]))/pow(h, 2)
```

These code fragments can later be inserted into corresponding templates to create compilable source code for PDE solvers. It is worth mentioning that in higher dimension and schemes of higher order of accuracy, such kernels become very complex to derive and implement manually.

## 3.3 Code generation

Once we have generated the mathematical kernel with SymPy, we are able to insert the code fragments into pre-prepared template files to create compilable source code. There are several Python libraries for web development, such as Jinja2[1], django[2] and Mako[3], which provide sufficiently sophisticated template facility for our purposes. We note that it is sufficient for the aim of OPESCI-FD to prepare separate template files for each set of PDEs and numerical schemes. However, efficiency and scalability can be improved by exploiting the common structure of such problems and create hierarchy of templates that share common modules, with possible application of Python meta-programming functionality.

---

[1] http://jinja.pocoo.org/docs/dev/
[2] https://www.djangoproject.com/
[3] http://www.makotemplates.org/

In this chapter we have demonstrated the basic techniques used by OPESCI-FD to generate the computational kernels. Next we will discuss the typical FD schemes used in seismic modelling. To be able to generate code which implements these schemes is the initial target of OPESCI-FD.

# Chapter 4

# FD schemes for 1D seismic wave equation

In this chapter we introduce common FD schemes used in seismic studies. This sets up the scope of possible schemes that OPESCI-FD can explore. We start with 1D problems to study the essential concepts and approaches without the unnecessary complexity of 2D or 3D problems. The common themes shared across these schemes strongly influenced the design of OPESCI-FD. This chapter closely follows the discussion in [42].

## 4.1 Equation of motion and the stress-strain relationship

Consider a perfectly elastic isotropic medium in 1D with density $\rho(x)$ and elastic Lamé coefficients $\mu(x)$ and $\lambda(x)$, where $\rho(x)$, $\mu(x)$ and $\lambda(x)$ are continuous functions of the spatial coordinate $x$. We further denote the displacement vector as $u(x,t)$ and stress tensor as $\sigma_{ij}(x,t)$ respectively, both being continuous function of $x$ and time $t$. This fully defines a 1D problem along the $x$-axis.

Newton's Law and Hooke's Law govern the equations of motion and the stress-strain relationship. We use $M$ as the corresponding elastic modulus. This relationship can be written in one of the following equivalent formulations as differential equation or set of simultaneous differential equations (here we omit the net body-force term). Also note that different wave forms (longitudinal or transverse) can be expressed with the same equation(s). Table 4.1 shows the configuration of waves propagating in $x$ direction, with their corresponding elastic moduli and formal names in seismic studies.

| Displacement $u$ | Stress tensor $\sigma$ | Elastic modulus $M$ | Wave name |
| --- | --- | --- | --- |
| $u_x$ | $\sigma_{xx}$ | $\lambda + 2\mu$ | P wave |
| $u_y$ | $\sigma_{xy}$ | $\mu$ | SH wave |
| $u_z$ | $\sigma_{xz}$ | $\mu$ | SV wave |

Table 4.1: Configuration of waves propagating in $x$ direction

***Displacement formulation***

$$\rho \frac{\partial^2 u}{\partial t^2} = \frac{\partial}{\partial x}\left(M \frac{\partial u}{\partial x}\right). \tag{4.1}$$

In this formulation the only dependent variable is the displacement $u$. Note that in the case of a homogeneous medium, the elastic modulus can be taken out of the differentiation and the equation simplifies to

$$\rho \frac{\partial^2 u}{\partial t^2} = M \frac{\partial^2 u}{\partial x^2}. \tag{4.2}$$

The speed of wave in the medium $c$ is given by the relation $c^2 = M/\rho$, and equation (4.2) can be written as

$$\frac{\partial^2 u}{\partial t^2} = c^2 \frac{\partial^2 u}{\partial x^2}. \tag{4.3}$$

This is the familiar 1D acoustic wave equation that we discussed in Chapter 3. One advantage of displacement formulation is that only one field $u$ needs to be modelled. Certain stencil tools, such as Pochoir, exploit this to build clean abstractions to describe the stencil topology. This abstraction partly restricts application on other formulations discussed in the following section.

***Displacement-stress formulation***

$$\rho \frac{\partial^2 u}{\partial t^2} = \frac{\partial \sigma}{\partial x}, \tag{4.4}$$

$$\sigma = M \frac{\partial u}{\partial x}. \tag{4.5}$$

In this formulation we also need to model stress $\sigma$. Note that this scheme only move forward in time domain in Equation (4.4), while Equation (4.5) is a transformation in the same time step.

***Displacement-velocity-stress formulation***

$$\rho \frac{\partial v}{\partial t} = \frac{\partial \sigma}{\partial x}, \tag{4.6}$$

$$v = \frac{\partial u}{\partial t}, \tag{4.7}$$

$$\sigma = M \frac{\partial u}{\partial x}. \tag{4.8}$$

Here we introduce velocity function $v = \partial u/\partial t$, and rewrite (4.4) into (4.6). This lower the second order derivative $\partial^2 u/\partial t^2$ into first order derivative $\partial v/\partial t$. The scheme move forward in time domain in both Equations (4.6) and (4.7). Equation (4.8) again is a transformation in the same time step.

***Velocity-stress formulation***

$$\rho \frac{\partial v}{\partial t} = \frac{\partial \sigma}{\partial x}, \tag{4.9}$$

$$\frac{\partial \sigma}{\partial t} = M \frac{\partial v}{\partial x}. \tag{4.10}$$

Here Equation (4.10) is derived by differentiating (4.8) with respect to time . Compared to other formulations, the velocity-stress formulation has the advantage of exhibiting desirable symmetry between the velocity field and the stress field. Only first order derivatives in space and time exist in the PDEs. Further more, the velocity field and stress field are decoupled, so that they can be updated independently from each other. This symmetry can result in significant advantage in memory utilisation of the time updating process [22].

## 4.2 FD schemes

In this section we discuss the common FD schemes used in seismic wave modelling .We will use the tuple *(spatial accuracy, temporal accuracy)* to denote the accuracy of a FD scheme. As an example, in a (2,2) scheme, second-order approximation will be used for both spatial and temporal derivatives.

### 4.2.1 1D displacement (2,2) scheme on regular grid

Displacement schemes are normally implemented on regular (non-staggered) structured grids. Consider a conventional structured discrete space-time grid similar to Figure 3.1, with time step $\Delta t$ and grid spacing $h$. We use the second-order centred FD approximation to replace both the spatial and temporal second derivatives in Equation (4.3) to obtain the difference equation

$$\frac{1}{\Delta t^2}(u_i^{m+1} - 2u_i^{m-1} + u_i^{m-1}) = c^2 \frac{1}{h^2}(u_{i+1}^m - 2u_i^m + u_{i-1}^m). \qquad (4.11)$$

This leads to an explicit FD scheme:

$$u_i^{m+1} = 2u_i^m - u_i^{m-1} + (c\frac{\Delta t}{h})^2(u_{i+1}^m - 2u_i^m + u_{i-1}^m). \qquad (4.12)$$

The truncation error $TE$ of the FD approximation is given by the difference between the difference equation (4.11) and the Taylor expansion of original differential equation (4.3)

$$TE = \frac{1}{12}(\frac{\partial^4 u}{\partial t^4}\Delta t^2 - c^2\frac{\partial^4 u}{\partial x^4}h^2) + O(\Delta t^4) + O(h^4). \qquad (4.13)$$

The leading term of the truncation error is consistent with the scheme having second-order accuracy in both space and time. We therefore use (2,2) to describe the accuracy of the scheme.

## 4.2.2 1D displacement-stress (2,2) scheme on spatially staggered grid

Displacement-stress schemes are commonly implemented on a spatially staggered grid, where the displacement fields are stored on whole grid points, and the stress fields are stored on half grid points (or vice versa). A typical layout is illustrated in Figure 4.1. The staggered grid has an important desirable feature. When using second-order FD approximations, the first spatial derivative $\partial u/\partial x$ is calculated at the mid-points between two whole grid points, and these mid-points coincide with the locations of stress field $\sigma$ which is stored at half grid points. Therefore, $\partial^2 u/\partial t^2$ and $\partial \sigma/\partial x$ in Equation (4.4) are at the same position, which reduces errors. The same applies to $\sigma$ and $\partial u/\partial x$ in Equation (4.5).

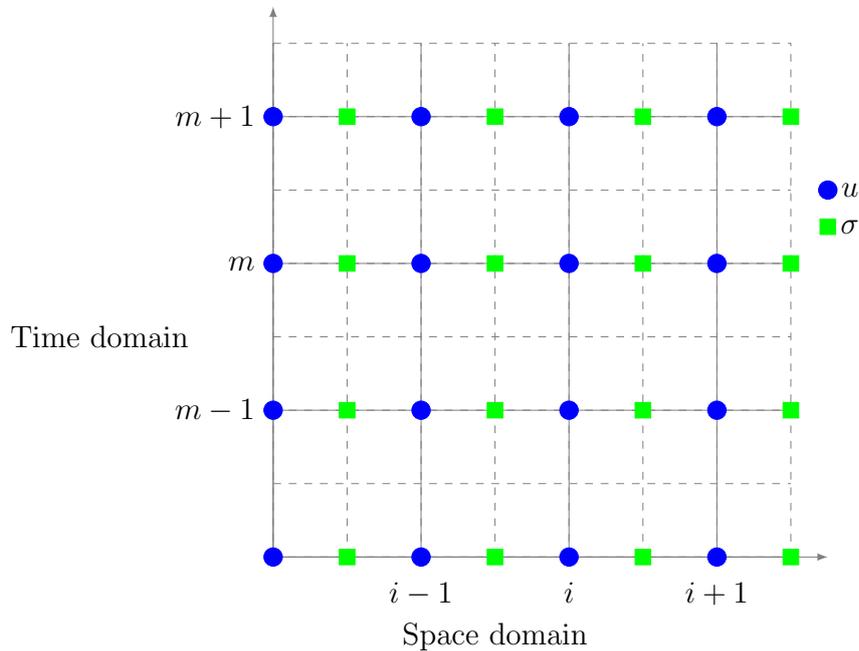

Figure 4.1: Spatially staggered grid for 1D displacement-stress scheme

Using second order FD approximations, we obtain the difference equation for $u$ at spatial location $i$ and time $m$ as

$$\rho_i \frac{1}{\Delta^2}(u_i^{m+1} - 2u_i^m + u_i^{m-1}) = \frac{1}{h}(\sigma_{i+\frac{1}{2}}^m - \sigma_{i-\frac{1}{2}}^m), \tag{4.14}$$

and the difference equation for $\sigma$ at spatial location $i + \frac{1}{2}$ and time $m$ is

$$\sigma_{i+\frac{1}{2}}^m = M_{i+\frac{1}{2}} \frac{1}{h}(u_{i+1}^m - u_i^m) \tag{4.15}$$

This lead to the following explicit time-stepping scheme

$$u_i^{m+1} = 2u_i^m - u_i^{m-1} + b_i \frac{\Delta t^2}{h}(\sigma_{i+\frac{1}{2}}^m - \sigma_{i-\frac{1}{2}}^m), \tag{4.16}$$

$$\sigma_{i+\frac{1}{2}}^m = M_{i+\frac{1}{2}} \frac{1}{h}(u_{i+1}^m - u_i^m), \tag{4.17}$$

where buoyancy, $b$, is defined as $1/\rho$. At each time step, we use (4.17) to calculate the stress field, then we can use the newly calculated stress field to update the displacement field for the next time step according to (4.16). However, if the stress field is not of interest to the application, we can substitute (4.17) into (4.16) and obtain the scheme

$$u_i^{m+1} = 2u_i^m - u_i^{m-1} + b_i (\frac{\Delta t}{h})^2 [M_{i+\frac{1}{2}}(u_{i+1}^m - u_{i-1}^m) - M_{i-\frac{1}{2}}(u_i^m - u_{i-1}^m)]. \tag{4.18}$$

This scheme only differs from the displacement scheme (4.12) by the formulation of effective elastic modulus. Apart from this, we see that the approximation of the equation in the displacement formulation on a conventional grid and the approximations of the equations in the displacement-stress formulation on a spatially staggered grid, in both cases using the second-order centred FD approximations to the temporal and spatial derivatives, led to the same explicit scheme, with the same error characteristics.

### 4.2.3 1D velocity-stress (2,2) scheme on fully staggered grid

For similar reasons as the displacement-stress formulation, velocity-stress schemes are commonly implemented with fully staggered grid. In a fully staggered grid in 1D, the stress field is shifted by half of the grid size in spatial direction, while the velocity field is shifted by half of the grid size in temporal direction. Such grid ensures approximations of $\partial v/\partial$ and $\partial \sigma/\partial x$ in (4.9) are centred at the same position, and the same applies to $\partial \sigma/\partial t$ and $\partial v/\partial x$ in (4.10). A typical layout is illustrated in Figure 4.2. Note that the stress field and velocity field can be swapped in their locations, this choice affects the implementation of boundary conditions.

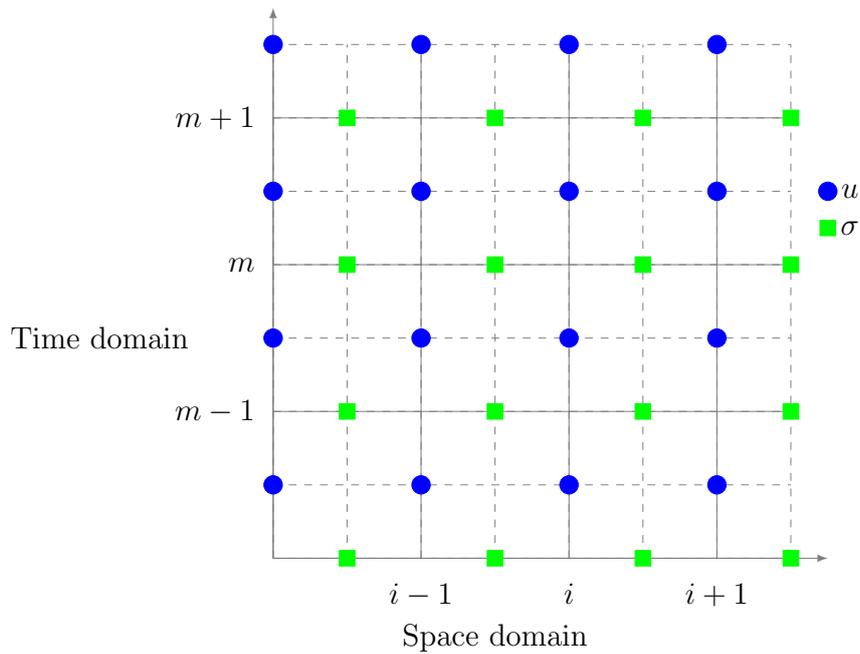

Figure 4.2: Fully staggered grid for 1D velocity-stress scheme

Again we use second-order FD approximations to replace the first spatial and temporal derivatives to obtain the 1D velocity-stress FD scheme

$$v_i^{m+\frac{1}{2}} = v_i^{m-\frac{1}{2}} + b_i \frac{\Delta t}{h}(\sigma_{i+\frac{1}{2}}^m - \sigma_{i-\frac{1}{2}}^m), \tag{4.19}$$

$$\sigma_{i-\frac{1}{2}}^m = \sigma_{i-\frac{1}{2}}^{m-1} + M_{i-\frac{1}{2}} \frac{\Delta t}{h}(v_i^{m-\frac{1}{2}} - v_{i-\frac{1}{2}}^{m-\frac{1}{2}}). \tag{4.20}$$

In each time step of the simulation, we first use Equation (4.20) to calculate the stress field explicitly. This advances the time by half a time step. Note that the right-hand-side of (4.20) only contains values at time $m-1$ and $m-1/2$, which are known at time $m$. Then we use Equation (4.19) to calculate the velocity field and advance the time by another half a time step. Similarly, the right-hand-side of (4.19) only contains values that are known at time $m+1/2$. This is an example of multistep methods known as two-step *leap-frog* scheme, where the fields are updated alternatively, and new values are calculated from values at two previous time steps. This scheme is also a (2,2) scheme, but we will improve the accuracy in next section.

### 4.2.4　1D velocity-stress (2,4) scheme on fully staggered grid

We can improve the accuracy of (2,2) velocity-stress scheme by applying higher order FD approximations to the first derivatives. Recall the governing PDEs of velocity-stress formulation:

$$\frac{\partial v}{\partial t} - b\frac{\partial \sigma}{\partial x} = 0, \tag{4.21}$$

$$\frac{\partial \sigma}{\partial t} - M\frac{\partial v}{\partial x} = 0. \tag{4.22}$$

The Taylor expansions of $v_i^{m+\frac{1}{2}}$ and $v_i^{m-\frac{1}{2}}$ with respect to $v_i^m$ are

$$\begin{aligned}v_i^{m+\frac{1}{2}} =& v_i^m + (\frac{\Delta t}{2})\frac{\partial v}{\partial t}\Big|_i^m + \frac{1}{2}(\frac{\Delta t}{2})^2\frac{\partial^2 v}{\partial t^2}\Big|_i^m, \\ &+ \frac{1}{6}(\frac{\Delta t}{2})^3\frac{\partial^3 v}{\partial t^3}\Big|_i^m + \frac{1}{24}(\frac{\Delta t}{2})^4\frac{\partial^4 v}{\partial t^4}\Big|_i^m + O(\Delta t^5)\end{aligned} \quad (4.23)$$

$$\begin{aligned}v_i^{m-\frac{1}{2}} =& v_i^m - (\frac{\Delta t}{2})\frac{\partial v}{\partial t}\Big|_i^m + \frac{1}{2}(\frac{\Delta t}{2})^2\frac{\partial^2 v}{\partial t^2}\Big|_i^m \\ &- \frac{1}{6}(\frac{\Delta t}{2})^3\frac{\partial^3 v}{\partial t^3}\Big|_i^m + \frac{1}{24}(\frac{\Delta t}{2})^4\frac{\partial^4 v}{\partial t^4}\Big|_i^m + O(\Delta t^5)\end{aligned} \quad (4.24)$$

Note that in this layout, velocity field is staggered in the time domain. As a result the location of $v_i^m$ is not on a whole grid point. By taking the difference of the Taylor expansions we obtain

$$v_i^{m+\frac{1}{2}} = v_i^{m-\frac{1}{2}} + \Delta t \frac{\partial v}{\partial t}\Big|_i^m + \frac{1}{24}\Delta t^3 \frac{\partial^3 v}{\partial t^3}\Big|_i^m + O(\Delta t^5) \quad (4.25)$$

Following similar procedures for the Taylor expansions of $\sigma_{i-\frac{1}{2}}^m$ and $\sigma_{i-\frac{1}{2}}^{m-1}$ with respect to $\sigma_{i-\frac{1}{2}}^{m-\frac{1}{2}}$, we can get

$$\sigma_{i-\frac{1}{2}}^m = \sigma_{i-\frac{1}{2}}^{m-1} + \Delta t \frac{\partial \sigma}{\partial t}\Big|_{i-\frac{1}{2}}^{m-\frac{1}{2}} + \frac{1}{24}\Delta t^3 \frac{\partial^3 \sigma}{\partial t^3}\Big|_{i-\frac{1}{2}}^{m-\frac{1}{2}} + O(\Delta t^5). \quad (4.26)$$

We then solve for the first temporal derivatives of $v$ and $\sigma$, and substitute back to the PDEs, while keeping the other terms the same

$$\frac{1}{\Delta t}(v_i^{m+\frac{1}{2}} - v_i^{m-\frac{1}{2}}) - \frac{1}{24}\Delta t^2 \frac{\partial^3 v}{\partial t^3}\Big|_i^m + O(\Delta t^4) - b\frac{\partial \sigma}{\partial x}\Big|_i^m = 0, \quad (4.27)$$

$$\frac{1}{\Delta t}(\sigma_{i-\frac{1}{2}}^m - \sigma_{i-\frac{1}{2}}^m) - \frac{1}{24}\Delta t^2 \frac{\partial^3 \sigma}{\partial t^3}\Big|_{i-\frac{1}{2}}^{m-\frac{1}{2}} + O(\Delta t^4) - M\frac{\partial v}{\partial x}\Big|_{i-\frac{1}{2}}^{m-\frac{1}{2}} = 0. \quad (4.28)$$

We use the fourth-order FD approximations for the first spatial derivatives, and truncate terms of $O(\Delta t^2)$ and above. We can obtain the explicit (2,4) velocity-stress scheme on a

staggered grid

$$v_i^{m+\frac{1}{2}} = v_i^{m-\frac{1}{2}} + b_i \frac{\Delta t}{h} [\frac{9}{8}(\sigma_{i+\frac{1}{2}}^m - \sigma_{i-\frac{1}{2}}^m) - \frac{1}{24}(\sigma_{i+\frac{3}{2}}^m - \sigma_{i-\frac{3}{2}}^m)], \qquad (4.29)$$

$$\sigma_{i-\frac{1}{2}}^m = \sigma_{i-\frac{1}{2}}^{m-1} + M_{i-\frac{1}{2}} \frac{\Delta t}{h} [\frac{9}{8}(v_i^{m-\frac{1}{2}} - v_{i-1}^{m-\frac{1}{2}}) - \frac{1}{24}(v_{i+1}^{m-\frac{1}{2}} - v_{i-2}^{m-\frac{1}{2}})]. \qquad (4.30)$$

We write the leading terms of the truncation error in velocity of this scheme, again by taking difference between the difference equation (4.27) and the Taylor expansion of the PDE (4.9)

$$TE = \frac{1}{24} \frac{\partial^3 v}{\partial t^3} \Delta t^2 + \frac{3}{640} b \frac{\partial^5 \sigma}{\partial x^5} h^4. \qquad (4.31)$$

The leading terms of the truncation error in stress is analogous. This confirms the scheme to have second-order accuracy in time, fourth-order accuracy in space.

The time stepping method is similar to (2,2) velocity-stress scheme. Stress field and velocity field are updated alternatively in each half time step.

### 4.2.5  1D velocity-stress (4,4) scheme on fully staggered grid

Lastly we look at the scheme that has fourth-order accuracy in both space and time. The leapfrog time update used in Section 4.2.4 is only second-order accurate in time, and it seems reasonable to use higher order methods for the time discretisation as well. However, the same method used for space discretisation to achieve fourth-order accuracy is not commonly applied directly to time domain. This is because in the resulting scheme, the coefficient of the most advanced time level would be significantly smaller in magnitude than the coefficients of the time level where the derivative is approximated. Therefore, the explicit equation for time update is likely to lead to numerical inaccuracy.[1] Further more, the increase in number of time levels needed in the kernels is undesirable, as it requires more computer memory.

In order to keep the $O(\Delta t^2)$ terms in Equations (4.27) and (4.28), we need to approximate their coefficients which are third derivatives in time. We follow the Lax-Wendroff [34] approach to express the temporal derivatives with spatial derivatives, by differentiating (4.9) twice with respect to time, and substitute with (4.10). In the case of homogeneous medium where $M$ and $b$ are both invariant in space, the equations simplify considerably to

$$\begin{aligned}
\frac{\partial^3 v}{\partial t^3} &= \frac{\partial}{\partial t}[\frac{\partial}{\partial t}(b\frac{\partial \sigma}{\partial x})] \\
&= b\frac{\partial}{\partial t}(\frac{\partial^2 \sigma}{\partial t \partial x}) \\
&= b\frac{\partial}{\partial t}(\frac{\partial^2 \sigma}{\partial x \partial t}) \\
&= b\frac{\partial}{\partial t}(M\frac{\partial^2 v}{\partial x^2}) \\
&= bM^2\frac{\partial^3 v}{\partial x^3},
\end{aligned} \quad (4.32)$$

and similarly for stress field

$$\frac{\partial^3 \sigma}{\partial t^3} = b^2 M \frac{\partial^3 \sigma}{\partial x^3}. \quad (4.33)$$

---

[1] Nonetheless, such schemes can be automatically generated by OPESCI-FD without extra development effort.

Substituting these expressions for the third derivatives in time into Equations (4.27) and (4.28), the governing PDEs can be approximated as

$$\frac{1}{\Delta t}(v_i^{m+\frac{1}{2}} - v_i^{m-\frac{1}{2}}) - \frac{1}{24}\Delta t^2 b^2 M \frac{\partial^3 \sigma}{\partial x^3}\bigg|_i^m - b\frac{\partial \sigma}{\partial x}\bigg|_i^m = 0, \quad (4.34)$$

$$\frac{1}{\Delta t}(\sigma_{i-\frac{1}{2}}^m - \sigma_{i-\frac{1}{2}}^{m-1}) - \frac{1}{24}\Delta t^2 b M^2 \frac{\partial^3 v}{\partial x^3}\bigg|_{i-\frac{1}{2}}^{m-\frac{1}{2}} - M\frac{\partial v}{\partial x}\bigg|_{i-\frac{1}{2}}^{m-\frac{1}{2}} = 0. \quad (4.35)$$

We can then apply fourth-order FD approximations to the third derivatives in space. The resulting scheme only contains two time levels and can be updated in similar way as (2,4) scheme.

We should note however the above formulation is only valid for homogeneous media, and is therefore limited in usage cases. For heterogeneous medium , the formulation is considerably more complicated which includes spatial derivatives of $b$ and $M$. The method is also somewhat cumbersome in multi-dimensions as it requires mixed derivatives as well as derivatives of the material parameters. We plan to investigate the feasibility of systematically generated such formulations through symbolic manipulation in future development of OPESCI-FD.

In this chapter we have discussed different FD schemes used in seismic modelling. Such schemes are the motivating examples for OPESCI-FD to implement automatically. Next we will describe the high level design of OPESCI-FD and show how different components of OPESCI-FD work together to generate FD solvers.

## Chapter 5

# Design of OPESCI-FD

In this chapter we describe the high level organisation of OPESCI-FD. The principal emphasis of the design approach is to extract common themes from different FD schemes, and maintaining sufficient flexibility at the same time. As of August 2015, OPESCI-FD has successfully implemented velocity-stress scheme on staggered grid. However, special attention is paid to ensure that the framework can be extended easily to incorporate other schemes, such as those discussed in Chapter 4.

## 5.1 Data structures and abstractions

OPESCI-FD creates the following data structures to facilitate the code generation work flow. This separation of functionalities allows modularisation of the process, where each part can be extended individually and weaved together to solve specific problems. Figure 5.1 shows the class structure of OPESCI-FD. We will describe each of the classes in more details in this section.

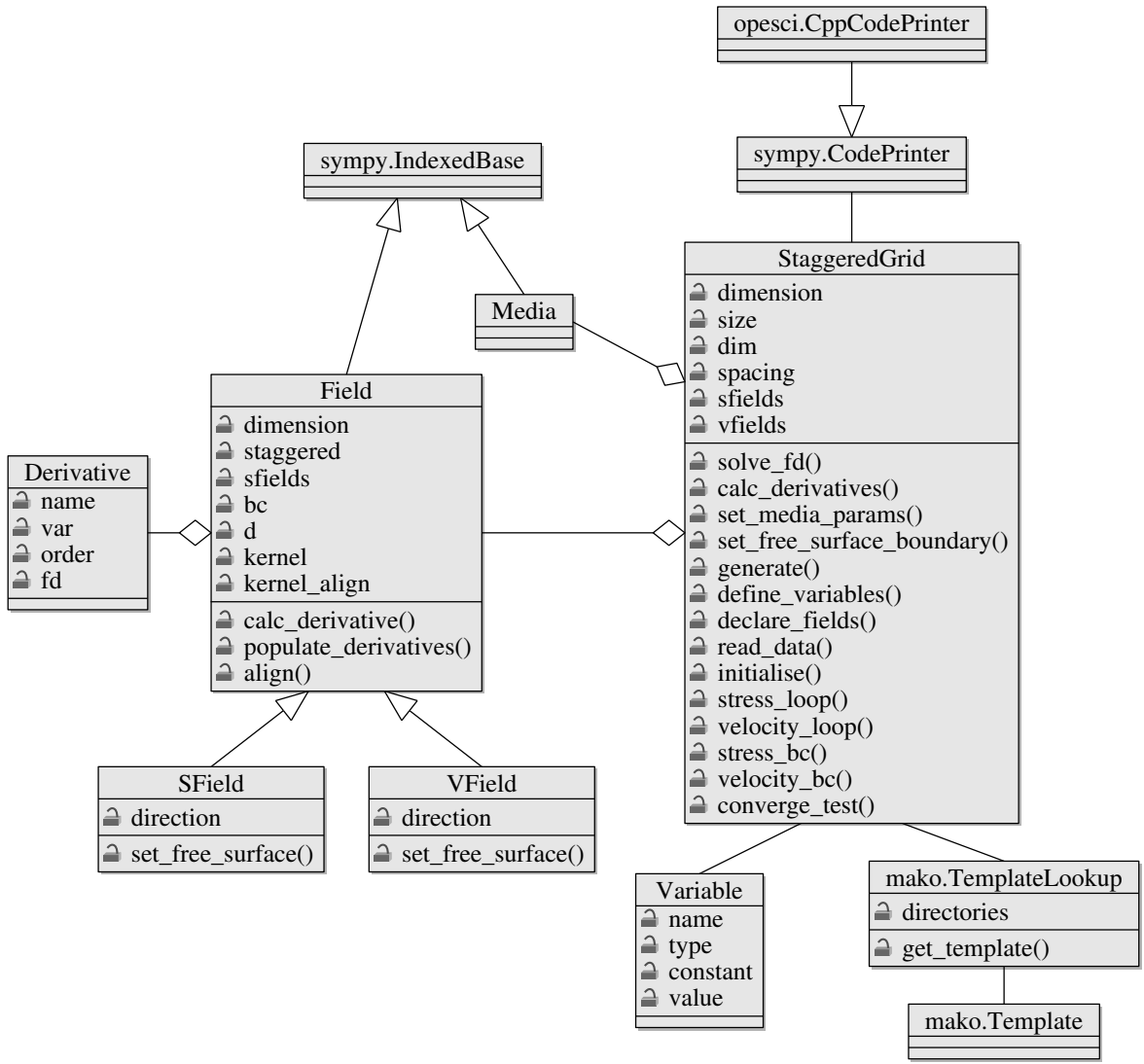

Figure 5.1: Class diagram of OPESCI-FD

#### *Field*

Field objects represent data stored on the grid, such as stress and velocity. It holds essential information about the field, such as its name and dimension. Because the scheme is on a staggered grid, the `staggered` attribute of the field (a tuple of boolean flags to indicate whether the field is staggered in a particular dimension) is particularly useful. The staggered information allows relative reference among field objects with ease. The `align` method uses the `staggered` attributes to facilitate converting field references into array indices in the generation of computational kernel. In some other stencil tools, it can be inconvenient to implement staggered grid based schemes, where the users have to mentally shift the references to integers. OPESCI-FD allows the users to work in the natural coordinate system of the staggered grids problems, without concerning about details such as array referencing.

Field class also implements operations such as calculation of FD approximations of derivatives. The `d` attribute is an nested list which stores the Derivative objects representing derivatives of the field of different order and different dependent variable. The `kernel` and `kernel_aligned` attributes store the mathematical kernel and its aligned version that would be used for time updating. The `bc` attribute stores the calculation for boundary grids. These data will be required in different phases of the code generation process.

Field class extends IndexedBase class in SymPy, so that `[ ]` operator can be used to address one data point, and the existing code printing rules for IndexedBase objects can be easily extended. Field class is parent class of VField and SField classes. VField represents velocity fields, and SField represents stress fields. The `staggered` attributes are calculated differently for VField and SField objects. VField and SField classes implements `set_free_surface()` method to generate the code which updates boundary cells for that surface.

The effective media parameters discussed in Section 6.2 are also implemented as Field objects. Since they only exist as data containers and no methods are required, we do not need to create a separate subclass for the media parameters in OPESCI-FD.

### *Derivative*

Derivative[1] class is essentially a data structure to hold the derivatives of the associated Field object. Here `var` and `order` denote the dependent variable and order of the derivative. The `fd` attribute is a list that stores the FD approximation of the derivative up to an accuracy level defined when the Derivative object is created. This abstraction allows the user to specify the governing PDEs easily in OPESCI-FD. It is also worth noting that in many cases, although the same set of PDEs are needed to compute the boundary conditions, typically FD approximations of lower accuracy order are applied at boundary (due to the absence of values outside the domain). With Derivative class, OPESCI-FD handles such situations easily without user intervention. The same PDEs can be used in both the computational kernels and the boundary conditions by expanding the Derivative objects to different orders of accuracy.

---

[1] There already exists a Derivative class in SymPy, so the Derivative class in OPESCI-FD is named DDerivative instead.

### *Variable*

Variable class extends Symbol class in SymPy and holds extra information about the mathematical symbol, such as its value and whether it should be declared as constant or not. Variable objects are used to represent variables (or constants) that are needed in the generated code. The StaggeredGrid object maintains a dictionary of such Variable objects and is capable of generating code to declare them as specified. These objects include grid dimensions (named `dim1,dim2,dim3` in our example), grid spacings (named `dx1, dx2, dx3` in our example), time step size (named `dt` in our example). In the case of homogeneous media, the media parameters are also created as Variable objects as these parameters are constant across the entire domain. They are typically assigned by the user rather than read from input files.

### *StaggeredGrid*

StaggeredGrid is the main class that implements most of the functionalities of OPESCI-FD. The user will first create a StaggeredGrid object, which can generate computational kernels as symbolic expressions from input PDEs using `solve_fd()` method. The StaggeredGrid class implements various methods to generate code fragments for different sections of the final code, such as `declare_field()` and `stress_loop()`. They return strings which can then be inserted into template files to form compilable source code. This last insertion step is wrapped up in `generate()`. The user can change settings of the StaggeredGrid object and call `generate()` to create multiple versions of source code with minor modifications. A detailed example of the work flow is described in Chapter 6.

StaggeredGrid class is the main point of user interaction where the user controls the characteristics of the code generation process. This is achieved through various boolean switches in the class, as listed in table 5.1. These switches allow subtle changes of the generated code, for example, by setting `expand` switch the user can easily choose to factorise or expand the mathematical kernel in the generated code and compare their performances. The functionality of these switches are implemented at the level of symbolic mathematics. This process of amending code is therefore more robust and thorough compared with other practices such as manual textual replacement.

Table 5.1: Switches implemented in StaggeredGrid class

| switch | functionality | default value |
| --- | --- | --- |
| omp | insert `#pragma omp for` before outer loops | True |
| ivdep | insert `#praga ivdep` before inner loop | True |
| simd | insert `#pragma simd` before inner loop | False |
| double | use `float` (False) or `double` (True) for real numbers | False |
| io | include header files for io (e.g. vtk support) | False |
| read | read media parameters from input file | False |
| expand | expand kernel fully (no factorisation) | True |
| eval_const | pre-evaluate all constants in kernel in generated code | True |

*CppCodePrinter*

SymPy provides a CodePrinter class which takes an expression and generates one line of C++ code as a string. It is essentially a mapping of rules to use to convert different symbolic objects to string. The library is implemented in a highly extensible way such that the user can easily subclass it and override certain methods to fine-tune the behaviour. OPESCI-FD creates the CppCodePrinter class and reimplemented various methods for our specific purposes, such as printing Field objects as multidimensional arrays (*e.g.* `U[t,x,y,z]` printed as `U[t][x][y][z]`), and printing floating point numbers in scientific notation. We also implement printing equations as assignment statements in C++ for convenience. In the design of OPESCI-FD, this functionality is independent from rest of the code. We envision that the user will only need to extend this part of the framework (in addition to creating suitable template files) in order to generate code for other programming languages, such as Fortran, or other stencil compilers, such as Pochoir and other polyhedral tools.

### *mako.TemplateLookup*

`mako.TemplateLookup` is a collection class for template files used in Mako library. The StaggeredGrid object will use the associated collection to search for the template files for code generation. A mako template is simply a text file[2] labelled with variables which can be substituted by passing in keyword arguments to `render()` or `render_context()` methods. In OPESCI-FD, all template variables are substituted with code fragments generated by the grid objects.

## 5.2 OPESCI-FD code generation work flow

In this section, we summarise the key steps to generate FD solver with OPESCI-FD. More in-depth discussion can be found in Chapter 6, where we describe the implementation details of OPESCI-FD with a real world problem as an example.

The general steps to use OPESCI-FD are as follows:

1. Create Grid object (named `grid` here), set the essential parameters such as dimension, domain size, time step size, simulation time.

2. Create Field objects, associate the Field objects to `grid`.

3. Call `grid.calc_derivatives()` to create the Derivative objects of the fields.

4. Write the PDEs using the Derivative objects created. Pass the PDEs to `grid`.

5. Call `grid.solve_fd()` to compute the mathematical kernels.

6. Set initial conditions of the Field objects.

---

[2]Technically, this text file is first compiled into a Python module when it is used.

7. Set boundary conditions using *e.g.* `grid.set_free_surface_boundary()`.

8. Call `grid.generate()` to create the source code.

One of the highlights of current development of OPESCI-FD is the internalisation of compilation and testing inside the Grid objects. This reduces the overhead in repeated testing and allows users to work on a single clean Python interface.

In this chapter we have described the key design features of OPESCI-FD. Next we will demonstrate using OPESCI-FD to solve a real world example, and in doing so, we discuss the implementation of OPESCI-FD classes in more details.

This page intentionally left blank.

# Chapter 6

# OPESCI-FD code generation for 3D staggered grid scheme

In this chapter, we apply OPESCI-FD to solve a real world problem, namely 3D seismic wave propagation, with velocity-stress FD scheme on staggered grid. We also discuss the implementation detail of OPESCI-FD using this example. We start with description of the different building blocks required by the problem, and show how the components of OPESCI-FD, as discussed in Chapter 5, work together to create these building blocks and produce a compilable C++ program to solve the seismic wave equations.

Velocity-stress scheme on staggered grid, as discussed in Section 4.2.4, was first introduced by Madariaga [40] in 1970s. Virieux described the 2D (2,2) scheme for modelling SH waves [54] and later, P-SV waves [55]. Levander [35] extended the scheme to fourth-order spatial accuracy. Graves [22] introduced the 3D version of the (2,4) scheme. The staggered-grid schemes remains widely used in the FD modelling of seismic wave propagation because of its robustness towards a wide range of $V_p/V_s$ ratio. Later researches focused on optimising the algorithmic performance and methodologies to handle non-planar boundaries and heterogeneous media. In this section we demonstrate the work flow of using OPESCI-FD to generate source code for a 3D (2,4) velocity-stress wave propagator, mostly following Graves's formulation. More details about the numerical accuracy and stability analysis of the scheme can be found in [35] and [48].

## 6.1 Problem description

We start with the set of first-order PDEs describing the velocity-stress formulation. In linear isotropic elastic media, the velocity components are given by the 3D equivalent of Equation (4.9)

$$\begin{aligned}
\frac{\partial v_x}{\partial t} &= b(\frac{\partial \sigma_{xx}}{\partial x} + \frac{\partial \sigma_{xy}}{\partial y} + \frac{\partial \sigma_{xz}}{\partial z}), \\
\frac{\partial v_y}{\partial t} &= b(\frac{\partial \sigma_{xy}}{\partial x} + \frac{\partial \sigma_{yy}}{\partial y} + \frac{\partial \sigma_{yz}}{\partial z}), \\
\frac{\partial v_z}{\partial t} &= b(\frac{\partial \sigma_{xz}}{\partial x} + \frac{\partial \sigma_{yz}}{\partial y} + \frac{\partial \sigma_{zz}}{\partial z}).
\end{aligned} \quad (6.1)$$

The stress components are given by the 3D equivalent of (4.10)

$$\begin{aligned}
\frac{\partial \sigma_{xx}}{\partial t} &= (\lambda + 2\mu)\frac{\partial v_x}{\partial x} + \lambda(\frac{\partial v_y}{\partial y} + \frac{\partial v_z}{\partial z}), \\
\frac{\partial \sigma_{yy}}{\partial t} &= (\lambda + 2\mu)\frac{\partial v_y}{\partial y} + \lambda(\frac{\partial v_x}{\partial x} + \frac{\partial v_z}{\partial z}), \\
\frac{\partial \sigma_{zz}}{\partial t} &= (\lambda + 2\mu)\frac{\partial v_z}{\partial z} + \lambda(\frac{\partial v_x}{\partial x} + \frac{\partial v_y}{\partial y}), \\
\frac{\partial \sigma_{xy}}{\partial t} &= \mu(\frac{\partial v_x}{\partial y} + \frac{\partial v_y}{\partial x}), \\
\frac{\partial \sigma_{xz}}{\partial t} &= \mu(\frac{\partial v_x}{\partial z} + \frac{\partial v_z}{\partial x}), \\
\frac{\partial \sigma_{yz}}{\partial t} &= \mu(\frac{\partial v_y}{\partial z} + \frac{\partial v_z}{\partial y}).
\end{aligned} \quad (6.2)$$

In the above equations, $v_x$, $v_y$, $v_z$ are the velocity components, $\sigma_{xx}$, $\sigma_{yy}$, $\sigma_{zz}$, $\sigma_{xy}$, $\sigma_{xz}$, $\sigma_{yx}$ are the stress components, $\lambda$ and $\mu$ are the Lamé coefficients, and $b$ is the buoyancy. We implement the scheme on a 3D staggered grid as illustrated in Figure 6.1. In particular, normal stress fields $\sigma_{xx}$, $\sigma_{yy}$, $\sigma_{zz}$ are stored at whole grid points (*i.e.* grid points with integer indices), shear stress fields $\sigma_{xy}$, $\sigma_{xz}$, $\sigma_{yz}$ are stored at staggered grid points where the indices corresponding to the two subscripts are shifted by half the grid spacing (*e.g.* $\sigma_{xy}$ stored as $\sigma_{xy_{i+\frac{1}{2},j+\frac{1}{2},k}}$). The velocity fields $v_x$, $v_y$, $v_z$ are stored at staggered grid points where the index corresponding to the velocity direction is shifted by half the grid spacing (e.g. $v_x$ stored as $v_{x_{x+\frac{1}{2},y,z}}$). As a result, the normal stress components share the same grid positions, while each the shear stress and velocity component has its own grid position. The temporal locations of either the velocity fields or the stress fields need to be shifted by half the time step size. We made the arbitrary choice of shifting the velocity fields, so that stress fields are stored as $\sigma^m$ and velocity fields as $v^{m+\frac{1}{2}}$.

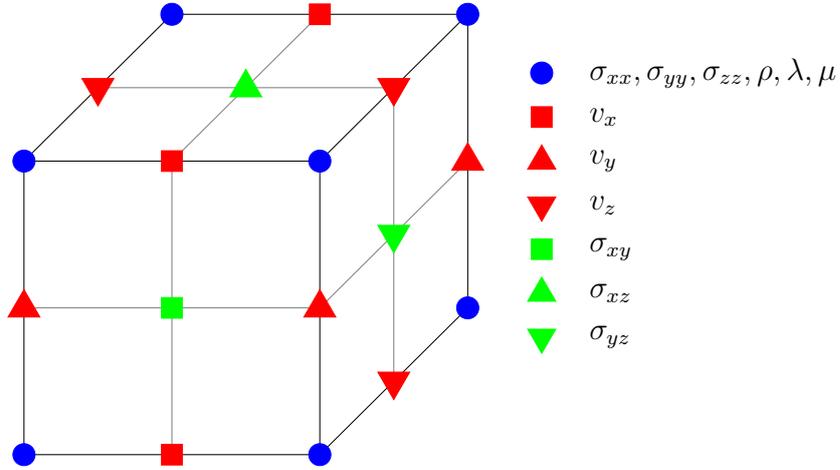

Figure 6.1: Unit cell for 3D staggered grid formulation.

Similar to the 1D case, the advantage of this formulation is that the various FD approximations applied to the PDEs are all naturally centred at the same point in space and time. This feature allows velocities to be updated independently from stresses at each grid point, thus exhibiting good parallelism characteristics and memory access pattern.

Using second-order FD approximations for the time derivatives, the updating equations for this scheme are

$$\begin{aligned}
v_{x_{i+\frac{1}{2},j,k}}^{m+\frac{1}{2}} &= v_{x_{i+\frac{1}{2},j,k}}^{m-\frac{1}{2}} + [b\Delta t(\frac{\partial \sigma_{xx}}{\partial x} + \frac{\partial \sigma_{xy}}{\partial y} + \frac{\partial \sigma_{xz}}{\partial z})]\Big|_{i+\frac{1}{2},j,k}^{m}, \\
v_{y_{i,j+\frac{1}{2},k}}^{m+\frac{1}{2}} &= v_{y_{i,j+\frac{1}{2},k}}^{m-\frac{1}{2}} + [b\Delta t(\frac{\partial \sigma_{xy}}{\partial x} + \frac{\partial \sigma_{yy}}{\partial y} + \frac{\partial \sigma_{yz}}{\partial z})]\Big|_{i,j+\frac{1}{2},k}^{m}, \\
v_{z_{i,j,k+\frac{1}{2}}}^{m+\frac{1}{2}} &= v_{x_{i,j,k+\frac{1}{2}}}^{m-\frac{1}{2}} + [b\Delta t(\frac{\partial \sigma_{xz}}{\partial x} + \frac{\partial \sigma_{yz}}{\partial y} + \frac{\partial \sigma_{zz}}{\partial z})]\Big|_{i,j,k+\frac{1}{2}}^{m},
\end{aligned} \quad (6.3)$$

for the velocity fields, and

$$
\begin{aligned}
\sigma_{xx}{}_{i,j,k}^{m+1} &= \sigma_{xx}{}_{i,j,k}^{m} + \Delta t \left[ (\lambda + 2\mu)\frac{\partial v_x}{\partial x} + \lambda(\frac{\partial v_y}{\partial y} + \frac{\partial v_z}{\partial z}) \right] \Big|_{i,j,k}^{m+\frac{1}{2}}, \\
\sigma_{yy}{}_{i,j,k}^{m+1} &= \sigma_{yy}{}_{i,j,k}^{m} + \Delta t \left[ (\lambda + 2\mu)\frac{\partial v_y}{\partial y} + \lambda(\frac{\partial v_x}{\partial x} + \frac{\partial v_z}{\partial z}) \right] \Big|_{i,j,k}^{m+\frac{1}{2}}, \\
\sigma_{zz}{}_{i,j,k}^{m+1} &= \sigma_{zz}{}_{i,j,k}^{m} + \Delta t \left[ (\lambda + 2\mu)\frac{\partial v_z}{\partial z} + \lambda(\frac{\partial v_x}{\partial x} + \frac{\partial v_y}{\partial y}) \right] \Big|_{i,j,k}^{m+\frac{1}{2}}, \\
\sigma_{xy}{}_{i+\frac{1}{2},j+\frac{1}{2},k}^{m+1} &= \sigma_{xy}{}_{i+\frac{1}{2},j+\frac{1}{2},k}^{m} + \Delta t \left[ \mu(\frac{\partial v_y}{\partial x} + \frac{\partial v_x}{\partial y}) \right] \Big|_{i+\frac{1}{2},j+\frac{1}{2},k}^{m+\frac{1}{2}}, \\
\sigma_{xz}{}_{i+\frac{1}{2},j,k+\frac{1}{2}}^{m+1} &= \sigma_{xz}{}_{i+\frac{1}{2},j,k+\frac{1}{2}}^{m} + \Delta t \left[ \mu(\frac{\partial v_z}{\partial x} + \frac{\partial v_x}{\partial z}) \right] \Big|_{i+\frac{1}{2},j,k+\frac{1}{2}}^{m+\frac{1}{2}}, \\
\sigma_{yz}{}_{i,j+\frac{1}{2},k+\frac{1}{2}}^{m+1} &= \sigma_{yz}{}_{i,j+\frac{1}{2},k+\frac{1}{2}}^{m} + \Delta t \left[ \mu(\frac{\partial v_z}{\partial y} + \frac{\partial v_y}{\partial z}) \right] \Big|_{i,j+\frac{1}{2},k+\frac{1}{2}}^{m+\frac{1}{2}},
\end{aligned}
\tag{6.4}
$$

for the stress fields. Here the subscripts refer to the spatial indices in the 3D grid, and the superscripts refer to the time index. The spatial derivatives are then substituted by four-order FD approximations

$$
\begin{aligned}
\frac{\partial \psi}{\partial x}\Big|_{i,j,k} &= \frac{1}{h_x}[\frac{9}{8}(\psi_{i+\frac{1}{2},j,k} - \psi_{i-\frac{1}{2},j,k}) - \frac{1}{24}(\psi_{i+\frac{3}{2},j,k} - \psi_{i-\frac{3}{2},j,k})], \\
\frac{\partial \psi}{\partial y}\Big|_{i,j,k} &= \frac{1}{h_y}[\frac{9}{8}(\psi_{i,j+\frac{1}{2},k} - \psi_{i,j-\frac{1}{2},k}) - \frac{1}{24}(\psi_{i,j+\frac{3}{2},k} - \psi_{i,j-\frac{3}{2},k})], \\
\frac{\partial \psi}{\partial z}\Big|_{i,j,k} &= \frac{1}{h_z}[\frac{9}{8}(\psi_{i,j,k+\frac{1}{2}} - \psi_{i,j,k-\frac{1}{2}}) - \frac{1}{24}(\psi_{i,j,k+\frac{3}{2}} - \psi_{i,j,k-\frac{3}{2}})],
\end{aligned}
\tag{6.5}
$$

where $\psi$ is the velocity or stress field, and $(h_x, h_y, h_z)$ are the 3D grid spacings. The leap-frog time updating method is as follow. Firstly the velocity fields at time $m + \frac{1}{2}$ are computed explicitly according to Equations (6.3) using the velocity fields at time $m - 1/2$ and stress fields at time $m$. Then the stress fields at time $m + 1$ are updated explicitly according to Equations (6.4) from the velocity fields just updated at time $m + 1/2$ together with the stress fields at time $m$. The cycle repeats for the next time step.

## 6.2 Media parameters

The media parameters $\rho$, $\lambda$, $\mu$, describe the physical properties of the elastic media. The differential operators in the PDEs act only on the fields, not on the media parameters, thus the media parameters can be taken out of the differentiation in this scheme. The complexity of the media does not affect the numerical formulation. The media parameters are stored at whole grid points with integer indices, sharing the locations with normal stresses, as shown in Figure 6.1. However, they are required at staggered grid points in calculating certain fields, as shown in Equation (6.3) and (6.4). For example, to compute $\sigma_{xy}$, $\mu$ is needed at location $(i+1/2, j+1/2, k)$ for integers $i$, $j$, $k$. This mismatch produces small errors in media with smooth variation, but in the cases of media with strong contrast, this could result in large errors and instability. To mitigate this, OPESCI-FD implements *effective media parameters* following the work of Randall [48] and Graves [22]. The effective media parameters essentially represent average values of physical parameters at regular (non-staggered) grid points depending on the directions applied. They are given by

$$\begin{aligned}
\bar{b}_x &= b(i+\tfrac{1}{2}, j, k) = \tfrac{1}{2}(b_{i,j,k} + b_{i+1,j,k}), \\
\bar{b}_y &= b(i, j+\tfrac{1}{2}, k) = \tfrac{1}{2}(b_{i,j,k} + b_{i,j+1,k}), \\
\bar{b}_z &= b(i, j, k+\tfrac{1}{2}) = \tfrac{1}{2}(b_{i,j,k} + b_{i,j,k+1}),
\end{aligned} \quad (6.6)$$

for the buoyancy, and

$$\begin{aligned}
\bar{\mu}_{xy} &= \mu(i+\frac{1}{2}, j+\frac{1}{2}, k) \\
&= [\frac{1}{4}(\frac{1}{\mu_{i,j,k}} + \frac{1}{\mu_{i+1,j,k}} + \frac{1}{\mu_{i,j+1,k}} + \frac{1}{\mu_{i+1,j+1,k}})]^{-1}, \\
\bar{\mu}_{xz} &= \mu(i+\frac{1}{2}, j, k+\frac{1}{2}) \\
&= [\frac{1}{4}(\frac{1}{\mu_{i,j,k}} + \frac{1}{\mu_{i+1,j,k}} + \frac{1}{\mu_{i,j,k+1}} + \frac{1}{\mu_{i+1,j,k+1}})]^{-1}, \\
\bar{\mu}_{yz} &= \mu(i, j+\frac{1}{2}, k+\frac{1}{2}) \\
&= [\frac{1}{4}(\frac{1}{\mu_{i,j,k}} + \frac{1}{\mu_{i,j+1,k}} + \frac{1}{\mu_{i,j,k+1}} + \frac{1}{\mu_{i,j+1,k+1}})]^{-1},
\end{aligned} \quad (6.7)$$

for the shear modulus (Lamé's second parameter). In the latter case, the average is taken on four points instead of two because shear stresses are staggered in two dimensions.

These effective media parameters replaces the corresponding parameters in equations (6.3) and (6.4). Note that first Lamé's first parameter $\lambda$ does not need to be replaced because the compression stresses are stored and calculated only at whole grid points.

## 6.3 Boundary conditions

From the mathematical kernels (6.3) and (6.4), we observe that for (2,4) schemes, the time update at one grid point normally requires values of other fields at coordinates $\pm 3/2$ grid sizes away at previous time step (although not all of other fields are needed). Therefore, different methods are needed to update the boundary grid points due to absence of values beyond the domain. The common techniques used are *imaging method* [35] and *adjusted FD approximations* [32]. In this section, we discuss the imaging method applied to free-surface boundary implemented in the current version of OPESCI-FD. The implementation largely follows the formulation in [22] and [42]. However, the flexibility of the framework enables smooth extension to other methods in the future.

Free-surface boundary conditions are normally used to model the Earth surface in seismic studies. The implementation of free-surface boundary is a challenging task because of issues related to numerical stability and accuracy of the reflected wave. As an example, consider a planar free surface of $x - y$ plane at $z = k$, with $z > k$ being out of the domain. The following *zero-stress* conditions need to be satisfied

$$\begin{aligned} \sigma_{xz} &= 0\big|_{z=k}, \\ \sigma_{yz} &= 0\big|_{z=k}, \\ \sigma_{zz} &= 0\big|_{z=k}. \end{aligned} \tag{6.8}$$

In (2,4) FD schemes, we need values at grid points up to two grid sizes beyond the boundary. The imaging method introduces two layers of *ghost cells* to ensure the zero-stress conditions. Once again, OPESCI-FD uses symbolic mathematics to derive and manipulate the equations at higher abstraction level and generate the corresponding code automatically.

To discretise the zero-stress formulation on the 3D staggered grid, we note that only $\sigma_{zz}$ field is stored at grid points located on the plane $z = k$, since $\sigma_{xz}$ and $\sigma_{yz}$ fields are staggered in $z$ dimension. The zero-stress conditions are therefore expressed through antisymmetry principle:

$$\sigma_{zz\,x,y,k} = 0, \qquad \sigma_{zz\,x,y,k-1} = -\sigma_{zz\,x,y,k+1}, \tag{6.9}$$

$$\sigma_{xz\,x,y,k-\frac{1}{2}} = -\sigma_{xz\,x,y,k+\frac{1}{2}}, \qquad \sigma_{zz\,x,y,k-\frac{3}{2}} = -\sigma_{xz\,x,y,k+\frac{3}{2}}, \tag{6.10}$$

$$\sigma_{yz\,x,y,k-\frac{1}{2}} = -\sigma_{yz\,x,y,k+\frac{1}{2}}, \qquad \sigma_{yz\,x,y,k-\frac{3}{2}} = -\sigma_{yz\,x,y,k+\frac{3}{2}}. \tag{6.11}$$

These equations immediately provide methods to update the stress fields of the ghost cells. Also note that $\sigma_{xx}$, $\sigma_{yy}$ and $\sigma_{xy}$ are not needed beyond the boundary in the computational kernels.

To compute the velocity fields at the ghost cells, we go back to the governing PDEs for stresses (6.4), but apply second-order FD approximations for the first spatial directives, instead of fourth-order approximations. These relationships, combined with Equations (6.9)-(6.11) and the values of $v_x$, $v_y$ and $v_z$ inside the domain are sufficient to calculate the values of velocity fields at the ghost cells. For example, to compute $v_z$ at the ghost cells, we use

$$\frac{\partial \sigma_{zz}}{\partial t}\bigg|_{z=k} = (\lambda + 2\mu)\frac{\partial v_z}{\partial z}\bigg|_{z=k} + \lambda\left(\frac{\partial v_x}{\partial x} + \frac{\partial v_y}{\partial y}\right)\bigg|_{z=k} = 0, \quad (6.12)$$

and the second-order FD approximations

$$\begin{aligned}
\frac{\partial v_x}{\partial x}\bigg|_{z=k} &= \frac{v_{x\,x+\frac{1}{2},y,k} - v_{z\,x-\frac{1}{2},y,k}}{h_x}, \\
\frac{\partial v_y}{\partial y}\bigg|_{z=k} &= \frac{v_{y\,x,y+\frac{1}{2},k} - v_{z\,x,y-\frac{1}{2},k}}{h_y}, \\
\frac{\partial v_z}{\partial z}\bigg|_{z=k} &= \frac{v_{z\,x,y,k+\frac{1}{2}} - v_{z\,x,y,k-\frac{1}{2}}}{h_z}.
\end{aligned} \quad (6.13)$$

After substituting (6.13) into (6.12), the only value outside the domain, and is therefore unknown after updating all the interior grids at one particular time step, is $v_{z\,x,y,k+1/2}$, which can be solved as

$$v_{z\,x,y,k+\frac{1}{2}} = v_{z\,x,y,k-\frac{1}{2}} + \frac{h_z\lambda}{h_x h_y(\lambda + 2\mu)}[h_x(v_{y\,x,y-\frac{1}{2},k} - v_{x\,x,y+\frac{1}{2},k}) \\ + h_y(v_{x\,x-\frac{1}{2},y,k} - v_{x\,x+\frac{1}{2},y,k})]. \quad (6.14)$$

We can apply the same procedure to other relationships in (6.9)-(6.11) to derive the expressions to compute $v_{x\,x,y,k+1}$ and $v_{y\,x,y,k+1}$ respectively.

Lastly, we observe that to compute $\sigma_{xx}$ and $\sigma_{yy}$ at the $z = k$ plane with the same kernels as the interior grid points, we need to approximate $\partial v_z/\partial z$ at $z = k$ to fourth-order accuracy. This requires value of $v_z$ at $z = k + 3/2$, which has not been calculated so far. A simple way around this is to use second-order approximation for the normal stresses at free-surface, or set velocity fields to zero at the outer ghost cells. This clearly results in loss of precision. Instead, we observe that from the PDE

$$\frac{\partial \sigma_{zz}}{\partial t} = (\lambda + 2\mu)\frac{\partial v_z}{\partial z} + \lambda\left(\frac{\partial v_x}{\partial x} + \frac{\partial v_y}{\partial y}\right) = 0 \quad (6.15)$$

at $z = k$, we can express $\partial v_z/\partial z$ with $\partial v_x/\partial x$ and $\partial v_y/\partial y$ as

$$\left.\frac{\partial v_z}{\partial z}\right|_{z=k} = \frac{-\lambda}{\lambda + 2\mu}\left[\frac{\partial v_x}{\partial x} + \frac{\partial v_y}{\partial y}\right]\bigg|_{z=k} \tag{6.16}$$

This relation holds regardless of the orders of approximation accuracy used. By substitute (6.16) into the governing PDEs (6.3), we eliminate the need for values of $v_z$ at $z = k + 3/2$. We can then derive new kernels to compute $\sigma_{xx}$ and $\sigma_{yy}$ at free-surface boundary $z = k$.

Implementation of boundary conditions in FDM is a challenging task. The methodologies are often logically clear, but operationally complicated. For the unit cube example described in Section 6.5, free-surface boundary conditions need to be applied to 9 fields on 6 sides. Manual implementation is prone to introduce subtle errors, such as off-by-one bugs, which can be hard to detect. By treating derivatives as a class and switch between second-order and fourth-order FD approximations, OPESCI-FD handles this naturally with manipulations and substitutions in symbolic mathematics. Once the methodology is decided, OPESCI-FD iterates systematically on different sides and fields, which greatly improves productivity and reduces errors.

## 6.4 Source description

In numerical seismic simulation, source terms are often needed in the framework to introduce perturbation into the media. In staggered grids, source implementation is explicit and can be achieved by simply adding the appropriate source components to the wave fields either as extra stress terms [14] or as extra velocity terms [58]. One of the most commonly used sources in seismic exploration is pressure point sources, such as buried explosives or offshore air-gun shots. In continuum mechanics, pressure $p$ and stress tensor $\boldsymbol{\sigma}$ are related by

$$\begin{aligned}p &= \frac{1}{3}tr(\boldsymbol{\sigma}) \\ &= \frac{1}{3}(\sigma_{xx} + \sigma_{yy} + \sigma_{zz})\end{aligned} \tag{6.17}$$

Usually we can treat explosive source as isotropic, *i.e.* $\sigma_{xx} = \sigma_{yy} = \sigma_{zz}$. The insertion of pressure source therefore only requires addition of another term to the normal stress fields. For example the modified computational kernel of $\sigma_{xx}$ becomes

$$\sigma_{xx}{}_{i,j,k}^{m+1} = \sigma_{xx}{}_{i,j,k}^{m} + \Delta t \left[ (\lambda + 2\mu)\frac{\partial v_x}{\partial x} + \lambda(\frac{\partial v_y}{\partial y} + \frac{\partial v_z}{\partial z}) \right] \Bigg|_{i,j,k}^{m+\frac{1}{2}} + s_{i,j,k}^{m+1}, \qquad (6.18)$$

where $s_{i,j,k}^{m+1}$ is the stress produced by the source at gird $(i,j,k)$ sampled at time $m+1$. In our implementation we keep the computational kernels unchanged in this case, instead we update the stress fields with the source terms after the updating loops at the end of each time step (this corresponds to `output_step` in the template, in Figure 6.12). This allows generating code on the same grid using different sources with minimum changes of the objects.

A similar procedure can be followed to add source terms to velocity fields. However, it is more complicated because the velocity fields are not defined at the same points in the staggered grid. This requires the insertion of velocity terms at two neighbouring points around the source location (*i.e.* as a dipole) for each velocity field.

Other sources in seismic problems can be more complicated and require more exact treatment of shear stress terms, which are ignored in the current implementation of OPESCI-FD.

## 6.5 Code verification

To study the accuracy of the implementation, OPESCI-FD implemented the propagation of 3D eigenwaves in a unit cube as a test case. Such test cases check numerical errors of the implemented scheme, and in doing so, makes rigorous software testing more trackable through continuous integration.

The (1, 1, 1) mode of the eigenwave in 3D is given by [17]

$$
\begin{aligned}
v_x &= \cos(\pi x)(\sin(\pi y) - \sin(\pi z))\cos(\Omega t), \\
v_y &= \cos(\pi y)(\sin(\pi z) - \sin(\pi x))\cos(\Omega t), \\
v_z &= \cos(\pi z)(\sin(\pi x) - \sin(\pi y))\cos(\Omega t), \\
\sigma_{xx} &= -A\sin(\pi x)(\sin(\pi y) - \sin(\pi z))\sin(\Omega t), \\
\sigma_{yy} &= -A\sin(\pi y)(\sin(\pi z) - \sin(\pi x))\sin(\Omega t), \\
\sigma_{zz} &= -A\sin(\pi z)(\sin(\pi x) - \sin(\pi y))\sin(\Omega t), \\
\sigma_{xy} &= 0 \\
\sigma_{xz} &= 0 \\
\sigma_{yz} &= 0
\end{aligned}
\quad (6.19)
$$

with $A = \sqrt{2\rho\mu}$ and $\Omega = \pi\sqrt{2\mu/\rho}$. The physical domain considered is a unit cube (1.0 × 1.0 × 1.0). The eigenwave functions provide the exact analytical solution of the fields at any given time, which we can compare with the numerical solutions at interior grid points to study the accuracy of the implementation. To do that, we first initialise the fields by setting $t = 0$ for stress eigenwave functions, and $t = \Delta t/2$ for velocity eigenwave functions. Then the simulation is run for $N$ time steps, and the approximate results at all grid points are compared with the exact solutions. The errors are quantified with the $L^2$ norms. For velocity fields the $L^2$ norm is given by

$$
L^2_{v_\alpha} = \sqrt{\int_{x,y,z}\left(v_\alpha\left(x,y,z,\left(N+\frac{1}{2}\right)\Delta t\right) - V_{\alpha x,y,z}^{N+\frac{1}{2}}\right)^2}, \quad (6.20)
$$

where $v_\alpha$ stands for the eigenwave functions for velocities $v_x$, $v_y$ and $v_z$, and $V_{\alpha x,y,z}$ is the numerical approximation of corresponding velocity fields at grid point $(x, y, z)$. The integration is taken over the entire domain, in practice we compute the difference at each interior grid point, and scale the result by the unit volume $dx \times dy \times dz$. Similarly the $L^2$ norm of

stress fields is

$$L^2_{\sigma_{\alpha\beta}} = \sqrt{\int_{x,y,z} \left(\sigma_{\alpha\beta}(x,y,z,N\Delta t) - T_{\alpha\beta}{}^N_{x,y,z}\right)^2}, \qquad (6.21)$$

where $\sigma_{\alpha\beta}$ represents the eigenwave functions for stress fields, and $T_{\alpha\beta_{x,y,z}}$ represents their numerical approximation at grid point $(x,y,z)$.

The symbolic mathematics framework makes generating code for calculating $L^2$ norms relatively straightforward. However, attention is needed at the details of some of the axillary C++ code, such as code for variable declaration and output to screen. Further more, we can study the convergence behaviour of the implemented scheme by calculating $L^2$ norms using different combination of $h$ and $\Delta t$. The result of applying this analysis on the example of unit cube eigenwave propagation is shown in Section 6.8.

Another way to check the correctness of the implementation is to use visualisation tools to visually inspect the evolution of the numerical solutions. OPESCI-FD uses the open-source VTK package[1] to create graphical output of the data. Users can set OPESCI-FD to generate code which saves down the values of chosen fields on the grid in VTK vector format. Projects with VTK support typically have long list of dependencies and are difficult to port to different systems. With that in mind, we have incorporated the build process in OPESCI-FD with CMake[2]. The generated source code can be easily compiled with CMake should the user choose to build with VTK output support. A current development of OPESCI-FD to incorporate the compilation process in Python is also under way. VTK images for the unit cube eigenwave example are shown in Section 6.8.

## 6.6 Code generation

The previous sections describe the essential building blocks of the OPESCI-FD framework. In this section we will demonstrate the entire code generation process with the 3D unit cube eigenwave propagation example. Below we show the abstracted version of the `eigenwave3d()` function which we created to generate C++ code implementing the 3D (4,2) FD scheme by linking together different components of OPESCI-FD. The generated code runs eigenwave simulation and outputs the computed $L^2$ norms of the fields. The complete code, with more elaborate documentation can, be found in OPESCI-FD project repository[3].

---

[1] http://www.vtk.org/
[2] http://www.cmake.org/

```
from opesci import *

def eigenwave3d(domain_size, grid_size, dt, tmax, o_step=False,
                o_converge=True, omp=True, simd=False, ivdep=True, io=False,
                double=False, filename='test.cpp', read=False, expand=True,
                eval_const=True, rho_file='', vp_file='', vs_file=''):
```

Figure 6.2: (4,2) FD scheme code generation function (part a).

Figure 6.2 shows the function signature of `eigenwave3d()`. Important parameters are `domain_size`, which is the size of the physical domain of the problem (for example, unit cube has domain size $(1.0, 1.0, 1.0)$); `grid_size` is a tuple that describes the number of grid points to use in each dimension; `dt`, the time step size; `tmax`, the time length to run the simulation; `o_step` the flag to switch on generating code to output the fields to disk at each time step (in VTK format); `o_converge`, the flag to switch on generating code which computes and outputs the $L2$ norms; and `filename` which specifies the location of the output source code. `rho_file`, `vp_file` and `vs_file` are paths to the physical media parameter files, they are not needed for this example. The other parameters are flags for StaggeredGrid object settings, and they have the same meaning as listed in Table 5.1. In this example, we can use the following code to generate source code to run the eigenwave simulation on the unit cube with 100 grid points in each dimension, for 1.0 second with time step of 0.01 second.

```
eigenwave3d((1.0,1.0,1.0), (100,100,100), 0.01, 1.0, o_step=False, o_converge=True,
            omp=True, simd=False, ivdep=True, io=False,
            filename=path.join(_test_dir, 'eigenwave3d.cpp'))
```

In Figure 6.3, lines 8 to 16 creates the SField and VField objects to represent the stress and velocity fields, and set their corresponding dimensions and directions. We use integers to represent directions, starting from 1 which stands for the $x$ direction. Velocity fields have one number as direction setting, e.g. direction of $v_x$ should be set to 1. Stress fields each require

---
[3]https://github.com/opesci/opesci-fd

```
 7   # Declare fields
 8   Txx = SField('Txx', dimension=3, direction=(1, 1))
 9   Tyy = SField('Tyy', dimension=3, direction=(2, 2))
10   Tzz = SField('Tzz', dimension=3, direction=(3, 3))
11   Txy = SField('Txy', dimension=3, direction=(1, 2))
12   Tyz = SField('Tyz', dimension=3, direction=(2, 3))
13   Txz = SField('Txz', dimension=3, direction=(1, 3))
14   U = VField('U', dimension=3, direction=1)
15   V = VField('V', dimension=3, direction=2)
16   W = VField('W', dimension=3, direction=3)
17
18   grid = StaggeredGrid(dimension=3, domain_size=domain_size,
19                        grid_size=grid_size,
20                        stress_fields=[Txx, Tyy, Tzz, Txy, Tyz, Txz],
21                        velocity_fields=[U, V, W])
```

Figure 6.3: (4,2) FD scheme code generation function (part b).

a pair of numbers as direction setting, *e.g.* direction of $\sigma_{xy}$ should be set to $(1,2)$. From the direction setting the objects are able to determine the grid layout of each fields, in particular the *staggered-ness* of the field in each direction using the rules described in Section 6.1. This is saved as the `staggered` attribute which is a tuple of boolean values in each Field object. Here we also made the arbitrary decision to stagger the time coordinates of velocity fields. For example $\sigma_{xz}$ will have `staggered=[False, True, False, True]` after this step. Line 18 to 21 then create a 3D StaggeredGrid object called `grid` with the parameters for domain size and grid size, and link the SField objects and VField objects with `grid`.

```
22   grid.set_accuracy([1, 2, 2, 2])
23
24   grid.set_time_step(dt, tmax)
25   grid.set_accuracy()
26   grid.set_switches(omp=omp, simd=simd, ivdep=ivdep, io=io,
27                     double=double, expand=expand,
28                     eval_const=eval_const)
29
30   # define parameters
31   rho, beta, lam, mu = symbols('rho beta lambda mu')
32   t, x, y, z = symbols('t x y z')
33   grid.set_index([x, y, z])
34
35   if read:
36       grid.set_media_params(read=True, rho_file=rho_file,
37                             vp_file=vp_file, vs_file=vs_file)
38   else:
39       grid.set_media_params(read=False, rho=1.0, vp=1.0, vs=0.5)
40
41   print 'require dt < ' + str(grid.get_time_step_limit())
```

Figure 6.4: (4,2) FD scheme code generation function (part c).

In Figure 6.4, line 22 sets the time step size and simulation time. Lines 24 to 26 set the switch of grid as specified by the user. Line 28 sets the accuracy order of the scheme, with the first number of the list as the temporal accuracy, followed by the spatial accuracy. Since only accuracy of even numbers can be implemented on staggered grid, we use half the actual accuracy order to express it in this input. In this example, we use second-order accuracy in time and fourth-order accuracy in space. The user can simply change this setting to generate code for other staggered grid schemes, for example, [1,4,4,4] specifies eighth-order accuracy in space. Caution is however required from the users to choose between schemes. The apparent scheme with fourth-order accuracy in space and time, set by [2, 2, 2, 2], is not stable in some circumstances [42]. The goal of OPESCI-FD is simply to implement the user specified FD schemes, and it is not a substitute to good mathematical analysis.

In this eigenwave example, the media parameters are constant across the whole domain, so it is not needed to read these values from files. Line 39 passes the values of the media parameters, namely density $\rho$, primary velocity $V_p$ and secondary velocity $V_s$ to `grid`. The grid will calculate buoyancy $b$, Lamé parameters $\lambda$ and $\mu$ using

$$
\begin{aligned}
b &= \frac{1}{\rho}, \\
\lambda &= \rho(V_p^2 - 2V_s^2), \\
\mu &= \rho V_s^2,
\end{aligned}
\quad (6.22)
$$

and create Variable objects for them. In the cases of reading the media parameters from files, `grid` will firstly generate code to call external functions to read the data from files into arrays, then generate nested loops to compute $b$, $\lambda$ and $\mu$, and finally generate the code to compute the effective media parameters according the formulation of Equations (6.6) and (6.7). Figure 6.5 shows part of the generated code to calculate the effective buoyancy in $x$ direction $b_x$, named `beta1` in the code. A typical implementation will have many such loops to read $\rho$, $V_p$, $V_s$, to calculate $b$, $\lambda$, $\mu$, and to calculate $b_x$, $b_y$, $b_z$, $\mu_{xy}$, $\mu_{xz}$, $\mu_{yz}$. The abstraction we have shown allows OPESCI-FD to generate such code systematically in a robust fashion. Using the maximum $V_p$ value and grid spacings $(dx, dy, dz)$, grid calculates the upper limit of time step size for stability and convergence. The user can choose `dt` based on this information.

```
// read rho from file
opesci_read_simple_binary_ptr("RHOx200",_rho_vec,8000000);
// calculate buoyancy
for(int x=2;x<dim1 - 2;++x){
for(int y=2;y<dim2 - 2;++y){
for(int z=2;z<dim3 - 2;++z){
beta[x][y][z]=1.0/rho[x][y][z];
}
}
}
// calculate effective buoyancy in x direction
for(int x=2;x<dim1 - 2;++x){
for(int y=2;y<dim2 - 2;++y){
for(int z=2;z<dim3 - 2;++z){
beta1[x][y][z]=5.0e-1*beta[x][y][z] + 5.0e-1*beta[x+1][y][z];
}
}
}
```

Figure 6.5: Generated code for media parameter calculation.

It is worth noting that OPESCI-FD generates all nested loops dynamically using a generic 1D loop template. For example, a nested loop of two levels is generated by first substituting the kernel into the generic 1D loop template, then this newly generated 1D loop is used as the kernel to create the 2D loop, using the same generic 1D loop template. This design approach allows the same StaggeredGrid implementation to be used for both 2D and 3D problems by simply changing the dimension parameter, therefore enhances reuse of the code base and eases maintenance burden.

In Figure 6.6, lines 42 to 53 define the eigenwave functions for each field, as described in equation (6.19). These eigenwave functions are the exact solution of the scheme, which we use to compare against the numerical solutions to compute errors. These functions exist in the form of symbolic mathematics objects (`Sympy.Expression`) which allows easy manipulation. For example, we can substitute $t, x, y, z$ with appropriate values to initialise the grid.

In Figure 6.7, line 63 computes the FD approximation of temporal and spatial derivatives systematically using techniques outlined in Chapter 3. This function populates the `d` attributes of all SField and VField objects linked with `grid` with newly created Derivative objects, which store the corresponding FD approximations as symbolic expressions. This

```
42  # define eigen waves
43  Omega = pi*sqrt(2*mu*beta)
44  A = sqrt(2*mu/beta)
45  U_func = cos(pi*x)*(sin(pi*y)-sin(pi*z))*cos(Omega*t)
46  V_func = cos(pi*y)*(sin(pi*z)-sin(pi*x))*cos(Omega*t)
47  W_func = cos(pi*z)*(sin(pi*x)-sin(pi*y))*cos(Omega*t)
48  Txx_func = -A*sin(pi*x)*(sin(pi*y)-sin(pi*z))*sin(Omega*t)
49  Tyy_func = -A*sin(pi*y)*(sin(pi*z)-sin(pi*x))*sin(Omega*t)
50  Tzz_func = -A*sin(pi*z)*(sin(pi*x)-sin(pi*y))*sin(Omega*t)
51  Txy_func = Float(0)
52  Tyz_func = Float(0)
53  Txz_func = Float(0)
54
55  U.set_analytic_solution(U_func)
56  V.set_analytic_solution(V_func)
57  W.set_analytic_solution(W_func)
58  Txx.set_analytic_solution(Txx_func)
59  Tyy.set_analytic_solution(Tyy_func)
60  Tzz.set_analytic_solution(Tzz_func)
61  Txy.set_analytic_solution(Txy_func)
62  Tyz.set_analytic_solution(Tyz_func)
63  Txz.set_analytic_solution(Txz_func)
```

Figure 6.6: (4,2) FD scheme code generation function (part d).

```
64  grid.calc_derivatives()
65
66  # PDEs: momentum equations
67  eq1 = Eq(U.d[0][1], beta*(Txx.d[1][1] + Txy.d[2][1] + Txz.d[3][1]))
68  eq2 = Eq(V.d[0][1], beta*(Txy.d[1][1] + Tyy.d[2][1] + Tyz.d[3][1]))
69  eq3 = Eq(W.d[0][1], beta*(Txz.d[1][1] + Tyz.d[2][1] + Tzz.d[3][1]))
70  # PDEs: stress-strain equations
71  eq4 = Eq(Txx.d[0][1], (lam + 2*mu)*U.d[1][1] + lam*(V.d[2][1]+W.d[3][1]))
72  eq5 = Eq(Tyy.d[0][1], (lam + 2*mu)*V.d[2][1] + lam*(U.d[1][1]+W.d[3][1]))
73  eq6 = Eq(Tzz.d[0][1], (lam + 2*mu)*W.d[3][1] + lam*(U.d[1][1]+V.d[2][1]))
74  eq7 = Eq(Txy.d[0][1], mu*(U.d[2][1] + V.d[1][1]))
75  eq8 = Eq(Tyz.d[0][1], mu*(V.d[3][1] + W.d[2][1]))
76  eq9 = Eq(Txz.d[0][1], mu*(U.d[3][1] + W.d[1][1]))
77
78  grid.solve_fd([eq1, eq2, eq3, eq4, eq5, eq6, eq7, eq8, eq9])
```

Figure 6.7: (4,2) FD scheme code generation function (part e).

step uses the order of accuracy from 2 up to the accuracy specified in `grid`. We use the first list index of `d` to indicate the dependent variable, and the second list index the order of the derivative. For example, `Txx.d[0][1]` stores the Derivative object representing $\partial \sigma_{xx}/\partial t$, so that `Txx.d[0][1].fd[2]` stores the expression for the fourth-order FD approximation of $\partial \sigma_{xx}/\partial t$. Similarly, `U.d[1][1].fd[1]` stores the second-order approximation for $\partial v_x/\partial x$. Note that in velocity-stress schemes, only first directives in time and space are needed in the PDEs, therefore, higher directives are not calculated and stored by `grid` in this case. However, the idea can be easily extended to work with arbitrary PDEs, since the underlying approximation technique can be used on directives of any order.

Line 69 to 78 are direct translation of PDEs (6.1) and (6.1), which create symbolic equations using the Derivative objects created in the previous step. Line 78 passes these equations to `grid`, which substitutes Derivative objects with the corresponding FD approximations. Internally, `grid` then creates simultaneous equations to solve for the computational kernel of each field for time update. This results in expressions similar to equations (6.3) and (6.4). The `grid` object also performs other tasks such as substitute the correct effective media parameters for each equation (in the case of reading data from files), and align the relative references (*i.e.* correctly resolving half indices in staggered grid) in the kernels so that they can be correctly transformed into array indices when converting to C++ code. The original and aligned version of the kernel are saved as attributes `kernel` and `kernel_aligned` of the Field objects. For example, the kernel of $\sigma_{xx}$ is shown in Figure 6.8, using `opesci.CppCodePrinter` to print out as C++ code.

```
>>> ccode(Txx.kernel_align)
(1.0F/24.0F)*(lambda*dt*dx1*dx2*(27*W[t0][x][y][z] + W[t0][x][y][z - 2] - 27*W[t0][x][y][z -
    1] - W[t0][x][y][z + 1]) + lambda*dt*dx1*dx3*(27*V[t0][x][y][z] + V[t0][x][y - 2][z] -
    27*V[t0][x][y - 1][z] - V[t0][x][y + 1][z]) + dt*dx2*dx3*(27*lambda*U[t0][x][y][z] +
    lambda*U[t0][x - 2][y][z] - 27*lambda*U[t0][x - 1][y][z] - lambda*U[t0][x + 1][y][z] +
    54*mu*U[t0][x][y][z] + 2*mu*U[t0][x - 2][y][z] - 54*mu*U[t0][x - 1][y][z] - 2*mu*U[t0][x
    + 1][y][z]) + 24*dx1*dx2*dx3*Txx[t0][x][y][z])/(dx1*dx2*dx3)
```

Figure 6.8: $\sigma_{xx}$ kernel (aligned).

It is worth noting that as the order of accuracy in time and space increases, the numerical method becomes more efficient and accurate. However, in (2,8) schemes for instance, such kernels becomes considerably more complicated for programmers to write and maintain. With the abstraction built by OPESCI-FD, they can be generated automatically for all nine fields with relative ease. OPESCI-FD also provides the options to expand the kernel and pre-value all the constants in it. The modified kernel of $\sigma_{xx}$ is shown in Figure 6.9. This reduces the readability but has the advantage of creating a more compact kernel, which could be easier for the compilers to reason about.

The last setting required in this example is to set the boundary conditions, shown in Figure 6.10. Here `dimension=1,side=0` sets the bottom surface in $x$ direction, i.e. the

```
>>> ccode(Txx.kernel_align)
Txx[t][x][y][z] + 5.625e-1*U[t][x][y][z] + 2.08333333333333e-2*U[t][x - 2][y][z] - 5.625e-1*
    U[t][x - 1][y][z] - 2.08333333333333e-2*U[t][x + 1][y][z] + 2.8125e-1*V[t][x][y][z] +
    1.04166666666667e-2*V[t][x][y - 2][z] - 2.8125e-1*V[t][x][y - 1][z] - 1.04166666666667e
    -2*V[t][x][y + 1][z] + 2.8125e-1*W[t][x][y][z] + 1.04166666666667e-2*W[t][x][y][z - 2] -
     2.8125e-1*W[t][x][y][z - 1] - 1.04166666666667e-2*W[t][x][y][z + 1]
```

Figure 6.9: $\sigma_{xx}$ kernel (aligned, with constants pre-valued).

```
79  # boundary conditions
80  grid.set_free_surface_boundary(dimension=1, side=0)
81  grid.set_free_surface_boundary(dimension=1, side=1)
82  grid.set_free_surface_boundary(dimension=2, side=0)
83  grid.set_free_surface_boundary(dimension=2, side=1)
84  grid.set_free_surface_boundary(dimension=3, side=0)
85  grid.set_free_surface_boundary(dimension=3, side=1)
```

Figure 6.10: (4,2) FD scheme code generation function (part f).

surface $x = 0$ in the unit cube, as free surface. Under the cover, `grid` will iterate through the stress and velocity fields, and apply the methodology described in section 6.3 to generate the code to update the ghost cells for that surface. The resultant code is saved as the `bc` attribute of the SField and VField objects, which is a nested list with the first index representing the dimension, second index representing the side of a particular surface. For example, `U.bc[1][0]` stores the code to update the ghost cells for $v_x$ at the boundary surface $x = 0$, as shown in Figure 6.11. Note that two layers of ghost cells have been added by `grid` on each surface of the domain, so here the plane $x = 0$ coincides with the array items with index $x = 2$, while items with $x = 0$ and $x = 1$ represent ghost cells. Also note that because $v_x$ is staggered in $x$ direction, the $x$ coordinate of the array item `U[t][1][y][z]` is actually 1.5, which is the ghost cell that we need update. The same applies to all other staggered indices. Considering there are 6 surfaces and 9 fields in our unit cube example, OPESCI-FD significantly reduces the efforts to generate code to update the ghost cells.

```
>>> U.bc[1][0]
U[t][1][y][z] = (dx1*dx2*lambda*W[t][2][y][z] - dx1*dx2*lambda*W[t][2][y][z - 1] + dx1*dx3*
    lambda*V[t][2][y][z] - dx1*dx3*lambda*V[t][2][y - 1][z] + dx2*dx3*lambda*U[t][2][y][z] +
    2*dx2*dx3*mu*U[t][2][y][z])/(dx2*dx3*(lambda + 2*mu));\n
```

Figure 6.11: $v_x$ ghost cell update at plane $x = 0$.

The last step in code generation is to insert the code fragments available at this stage into a prepared template file. OPESCI-FD uses the facility provided by Mako[4] library to achieve this. Figure 6.12 shows the template file that OPESCI-FD uses for implementing FD on staggered grid. The well-defined structure of the FD schemes means the code can generally be divided into individual segments, resulting in concise and clear templates. String labels in the templates enclosed by ${...} are keywords that need to be substituted with the code generated by `grid`. We will explain their meanings next.

```
1   <%include file="copyright.txt"/>
2   #ifdef _MSC_VER
3   #define M_PI 3.14159265358979323846
4   #endif
5   <%include file="common_include.txt"/>
6   % if io==True:
7   <%include file="io_include.txt"/>
8   % endif
9   #include <cmath>
10  #include <cstdio>
11  #include <string>
12
13  int main(){
14
15  ${define_constants}
16  ${declare_fields}
17  #pragma omp parallel
18  {
19      ${initialise}
20      ${initialise_bc}
21      for(int _ti=0;_ti<ntsteps;_ti++){
22          ${time_stepping}
23          ${stress_loop}
24          ${stress_bc}
25          ${velocity_loop}
26          ${velocity_bc}
27          ${output_step}
28      } // end of time loop
29  } // end of parallel section
30  ${output_final}
31  return 0;
32  }
```

Figure 6.12: Mako template file for staggered grid FD

### define_constant

This section of the code declares the variables and constants needed for the scheme, and assign them with appropriate values. Such constants include number of grids in each dimension `dim1, dim2, dim3`, grid spacings `dx1, dx2, dx3`, time step size `dt`, number of time steps `ntsteps` etc. The names of these constants are generated automatically, but can be changed by the user to improve readability. This section also includes the media parameters if they are constant throughout the domain. It is substituted with code generated by `grid.define_constant()`.

---

[4]http://www.makotemplates.org/

### define_fields

This section declares the array variables, mainly the stress and velocity fields. The first dimension of the arrays is the time index, followed by the spatial indices. The size of time indices is determined by the number of previous time steps needed in the leap-frog update. For (2,4) scheme, the array dimensions are `[2][dimx][dimy][dimz]`. OPESCI-FD enforces memory alignment in this step to facilitate vectorisation. The alignment size is defaulted to the page size of the architecture, but can be changed by user using `grid.set_alignment()`. When reading the media parameters from files, this segment also includes the code to read in the values, declare the arrays for the media parameters, and compute the effective media parameters. Media parameters are constant in time, so the corresponding array dimensions are `[dimx][dimy][dimz]`. This label is substituted with code generated by `grid.declare_fields()`.

### initialise

This section initialises the fields with their initial conditions. In the eigenwave test example, this is done by substituting $t = 0$ into the analytical functions of the stress fields, and $t = \Delta t/2$ into the analytical functions of the velocity fields. Nested loops are generated to assign the values to the fields at interior grid points (with time index 0). The $x$, $y$, $z$ values in the analytical functions need to be calculated for each field from the spatial indices, grid the spacing sizes and the staggered nature of the field. This step applies *first-touch* trick on Non-uniform memory access (NUMA) architecture automatically. This label is substituted with code generated by `grid.initialise()`.

### initialise_bc

This section populates the ghost cells with their initial values. It is similar to the boundary update in the main time loop, the only difference is that the time indices are 0 here. In the cases of free-surface boundary, the normal stresses do not need to be re-calculated according to Section 6.3 at initialisation. It is substituted with code generated by `grid.initialise_bc()`.

### time_stepping

This section updates the values of time variables for each time step. Time variables are variables used to reference the time indices of the arrays in the time loop. For second-order time-accurate schemes, the time variables are named `t0` and `t1`, where array items with time index `t1` are updated for each time step using array items with time index `t0`. Initially we set `t1=1` and `t0=0`, so that the stress field arrays at $t = \Delta t$, such as `Txx[1][x][y][z]`, can be updated with values from `U[0][x][y][z]`, `V[0][x][y][z]`, `W[0][x][y][z]` and `Txx[0][x][y][z]`. In the next time step, time variables update to `t1=0` and `t0=1`. This step will therefore overwrite the values at `Txx[1][x][y][z]`. This is safe because they are not needed in future updates. Higher order schemes follow the same idea. For example, in fourth-order time accurate schemes, `grid` will create time variables `t0`, `t1`, `t2` and `t3`, with `t3` indicating the array items to be updated. This label is substituted with code generated by `grid.time_stepping()`.

### stress_loop

This section generates the code to update the stress field arrays using the computational kernels similar to Figure 6.8. OPESCI-FD reads the kernel expressions saved in the SField objects, substitute `t` symbols with the appropriate time variables, and convert the kernels into C++ code using `opesci.CppCodePrinter`. By changing settings of StaggeredGrid object, suitable C++ preprocessor directives (such as `#pragma ivdep` and `#pragma simd`) are inserted to ensure the kernel is vectorized and the loop runs in parallel. This label is substituted with code generated by `grid.stress_loop()`. If flag `eval_const` is set to True, all constants in the kernel are substituted with their numerical values and pre-evaluated.

### stress_bc

This section updates the ghost cells of stress field arrays at each time step. OPESCI-FD iterates through all stress fields and surfaces and creates nested loops for each case, using the value stored in the `bc` attributes of the Field objects. For certain fields, values outside of some surfaces are not need, and are filtered out in this step. Preprocessor directives are also inserted for these generated loops. This label is substituted with code generated by `grid.stress_bc()`.

### velocity_loop

This section generates the code to update the velocity field arrays using the computational kernels stored in VField objects. The implementation is similar to `stress_loop`. However, the velocity fields need to be updated with the newly computed stress field values, so the time indices of the stress fields in the kernel are substituted by time variable `t1` instead of `t0`. This label is substituted with code generated by `grid.velocity_loop()`.

### velocity_bc

This section updates the ghost cells of velocity field arrays at each time step. It is implemented similarly as `stress_bc` and generated by `grid.velocity_bc()`.

### output_step

This section of the code creates output at each time step. In the eigenwave test example, OPESCI-FD saves down the selected fields to disk in VTK format in this section. The code for inserting source terms can also be included here. This label is substituted with code generated by `grid.output_step()`.

### output_final

This section is the final output of the program after the main time loop. For example, it can be used to save down the final state of the fields in VTK format. In the eigenwave test example, OPESCI-FD generate code to compute the $L2$ norms of all the fields in this section, by calling `grid.converge()`. This label is substituted with code generated by `grid.output_final()`.

The generated C++ source code for eigenwave test can be compiled directly, with either GCC, Clang or Intel compilers. For IO functionalities such as VTK support and reading media parameter files, the code need to be linked with various libraries. OPESCI-FD provides CMake files to automate compilation in these situations. In the current development version of OPESCI-FD, certain types of compilation have been incorporated inside the StaggeredGrid class by calling `grid.compile()`. This will be released in the next version of OPESCI-FD.

## 6.7 Stencil compilers

So far we have limited our scope to generate source code in C/C++, and improve performance with OpenMP API[5]. One of the main goals in OPESCI-FD design is to ensure flexibility to generate code for different downstream stencil compilers. We only need to extend `sympy.CodePrinter` class and create new templates to target different languages, with minimum changes to the symbolic mathematics library of OPESCI-FD.

In an earlier prototype of OPESCI-FD, we have attempted to generate code for the Pochoir compiler [52] in order to test the cache-oblivious trapezoidal decomposition algorithm. While we were able to generate most of the code needed, our approach was limited by the abstraction designed in Pochoir when we tried to implement the boundary conditions. Figure 6.13 shows how to specify a periodic boundary in Pochoir. Here `fd_bv_3d` is the name of the boundary function, $a$ is the Pochoir_Array object that represents the field, `t` is the time index, `x, y, z` are the spatial indices. Whenever the computation accesses the coordinates outside the domain, the boundary function is called to return the value to be used. One limitation imposed is only one field (a `Pochoir_Array` object) can be passed into the boundary function. However, in our implementation of free surface boundary in velocity-stress schemes, the ghost cells of velocity fields need to be updated with stress fields and other velocity fields. This requirement is not easily fulfilled in the Pochoir abstraction.

---

[5]http://openmp.org/wp/

```
1  Pochoir_Boundary_3D(fd_bv_3D , a, t, z, y, x)
2  return a.get(t, mod(z, a.size (2)), mod(y, a.size (1)), mod(x, a.size (0)));
3  Pochoir_Boundary_End
```

Figure 6.13: Example of 3D boundary function in Pochoir

Essentially, the velocity-stress FD schemes are solving a coupled set of equations using leap-frog updating, while the Pochoir abstractions such as Pochoir_Shape and Pochoir_Kernel are more suitable for single equations of scalar field. Furthermore, the trapezoidal decomposition algorithm might not be suitable for the access patterns of alternating leap-frog updating. However, we plan to incorporate Pochoir in OPESCI-FD in future development to incorporate other PDEs. For example, the acoustic equations for seismic waves (displacement FD formulations) can be solved with Pochoir compiler.

Polyhedral models which optimise memory access pattern by tiling become popular in recent years. Polly[6] is one of such tools built on LLVM infrastructure. Polly is designed to work on LLVM-IR (intermediate representation) to LLVM-IR transformation, so theoretically no change in the C++ source code is required to apply Polly optimisations. We plan to support Polly as one of the build options in the next version of OPESCI-FD which internalises compilation.

## 6.8 Experimental evaluation

In this section we present the results of the eigenwave analysis. The eigenwaves are simulated on a unit cube as described previously, with physical parameters $\rho = 1.0$, $\lambda = 0.5$, $\mu = 0.25$.

We run the simulation with 4 combination of $(h, \Delta t)$, namely $h = 0.1, 0.05, 0.025, 0.0125$, $\Delta t = 0.04, 0.01, 0.0025, 0.000625$ respectively. Here for each subsequent combination, we use $h$ value which is half of the previous combination, and $\Delta t$ value which is a quarter of the previous combination. All simulation is run until $t = 5.0$. We then compute the $L^2$ norms for each simulation and plot the error against $h$ on logarithmic scale. According to the truncation error of (2,4) FD scheme, Equation (4.31), this should result in a straight line on logarithmic scale. Note that we have chosen the same spacing size on each dimension, *i.e.* $h_x = h_y = h_z = h$. Single-precision floating point numbers are used for the calculation.

---
[6]http://polly.llvm.org/

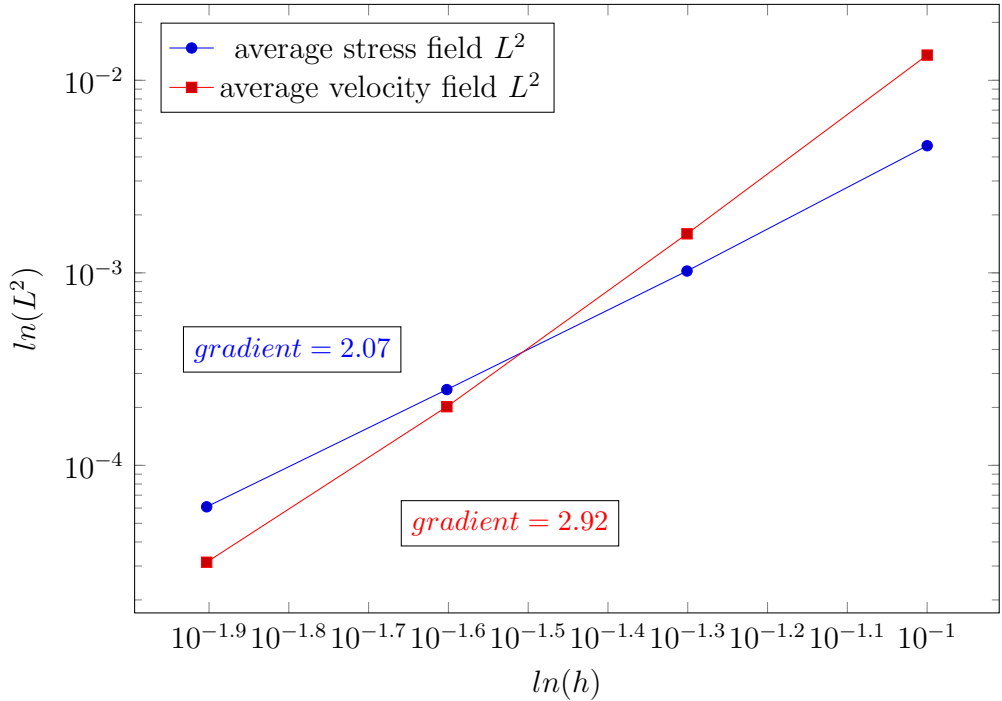

Figure 6.14: Convergence test for unit cube eigenwaves

The result is shown in Figure 6.14. We calculated the average $L^2$ norms of stress fields and velocity fields, because the individual $L^2$ norm of each field is very similar. Here we see the scheme converges linearly on logarithmic scale as expected. However, the rate of convergence is 2.07 for stress fields and 2.92 for velocity fields, both are less than 4 which is expected for (2,4) scheme. We hypothesise that this is primarily due to the reduction of accuracy at the free-surface boundary. Although stress fields are calculated with fourth-order accuracy in space, the velocity fields are reduced to second-order accuracy in space at the boundary, as shown in Equation (6.14). Also, the spatial variations introduced by the coefficients $\partial^3 v/\partial t^3$ and $\partial^5 \sigma/\partial x^5$ in the truncation errors might have an effect in the result. Future investigations, such as implementing and testing the (4,4) FD schemes, are needed to gain further insight into this convergence behaviour.

Visualisation of data can be extremely helpful in developing and checking FD implemen-tations. The VTK support built in OPESCI-FD makes exporting data for visualisation easy. This is a feature we use extensively in our own development and testing of OPESCI-FD. Figure 6.15 shows the evolution in time of $\sigma_{xx}$ on $x = 0.5$ plane in the unit cube eigenwave example. This corresponds to the exact solutions in Equation (6.19). The simulation is run with $h = 0.01$ and $\Delta t = 0.002$. The images are created with ParaView[7] using the data saved down by OPESCI-FD at each time step. Similarly, Figure 6.16 shows the evolution of $v_y$ on the same plane in the same eigenwave example. We observed that visual inspections of such graphs are very effective in detecting bugs in implementation of the scheme.

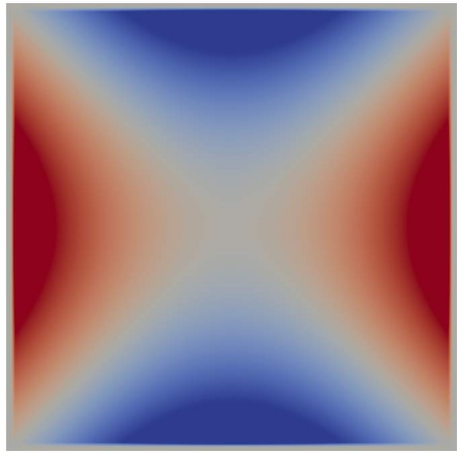 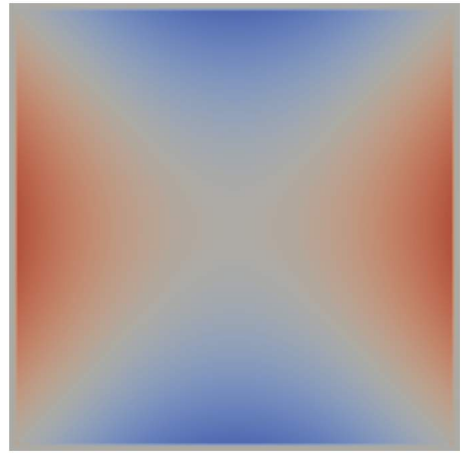

(a) $t = 0.5$        (b) $t = 1.2$

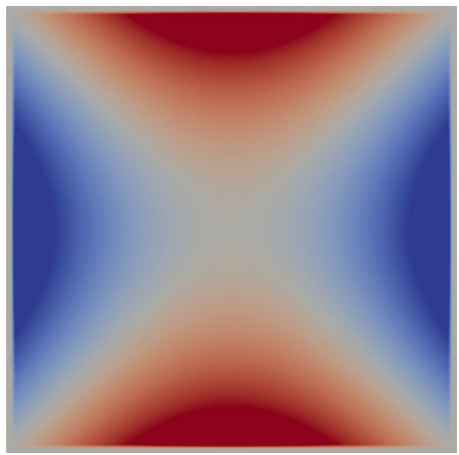 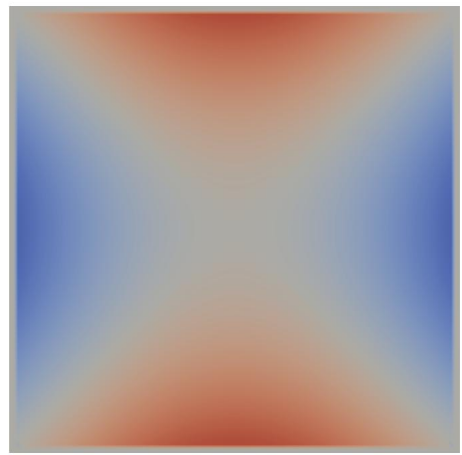

(c) $t = 1.9$        (d) $t = 2.6$

Figure 6.15: Evolution of $\sigma_{xx}$ on $x = 0.5$ plane in eigenwave example

---

[7] http://www.paraview.org/

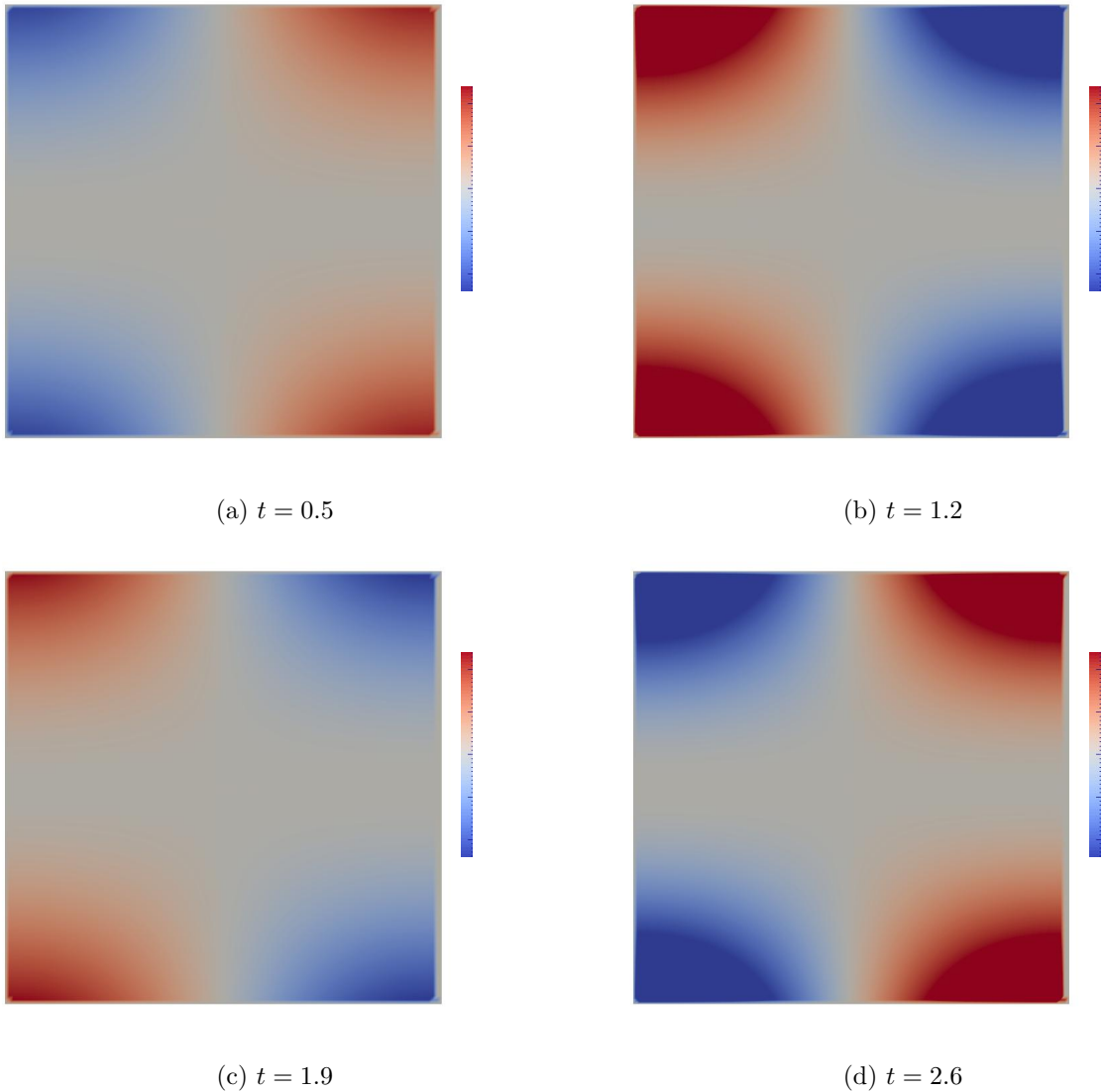

(a) $t = 0.5$      (b) $t = 1.2$

(c) $t = 1.9$      (d) $t = 2.6$

Figure 6.16: Evolution of $v_y$ on $x = 0.5$ plane in eigenwave example

The key concern in scientific computing is the performance of the code, and generating optimised code is a different challenge to generating code that simply works. To facilitate benchmarking, OPESCI-FD provides scripts to run benchmarks on the CX1 cluster[8], a HPC facility provided by Imperial College London.

The *roofline model* [27] is an intuitive performance model to gauge the limitations of numerical algorithms on specific architecture. Input to the model includes the architecture characteristics and the *Arithmetic Intensity* of the algorithm. We compiled and ran the generated code on Intel® Xeon® E5-2650 Sandy Bridge nodes on CX1. The node has clock frequency of 2.00GHz, 2 sockets with 8 cores on each socket, 256-bit wide AVX instructions (*i.e.* 8 single-precision floating point numbers), and can execute 2 vector instructors per clock cycle. The theoretical peak computation is therefore

$$\begin{aligned}
& (CPU\ Frequency) \times (\#Cores) \times (\#Instructions\ per\ Cycle) \\
& \times (\#Flops\ per\ Instruction) \\
& = 512\ GFlop/s
\end{aligned} \quad (6.23)$$

The maximum memory bandwidth is 51.2 GB/s per socket. The Arithmetic Intensity (AI) of an algorithm is derived from the number of floating point addition, multiplication, load and save instructions in the mathematical kernel as

$$AI = \frac{\#ADD + \#MUL}{(\#LOAD + \#STORE) \times word\ size}, \quad (6.24)$$

where `word size` is the number of floating point operands in each vector instruction (in this case `word size`=8). Because the peak flops is only achieved when two pipelines (one for addition, the other for multiplication) running simultaneously, we need to weight the AI to account for the imbalance between addition and multiplication instructions in the kernel. The weighted AI is given by

$$AI_{weighted} = AI \times \frac{\#ADD + \#MUL}{2 \times max(\#ADD, \#MUL)}. \quad (6.25)$$

OPESCI-FD provides functions to easily count the number of different types of instruc-tions in the kernel and calculate the AI. To avoid ambiguity, we generate the code with settings such that all factorisation is removed and all constants are pre-valued in the kernels. Finally, we calculate the combine AIs of the velocity kernel and stress kernel to get the overall weighted AI of the the (2,4) FD algorithm as 0.875, which corresponds to the theoretical peak performance of 89.6 Gflops/s.

---
[8]www.hpc.ic.ac.uk

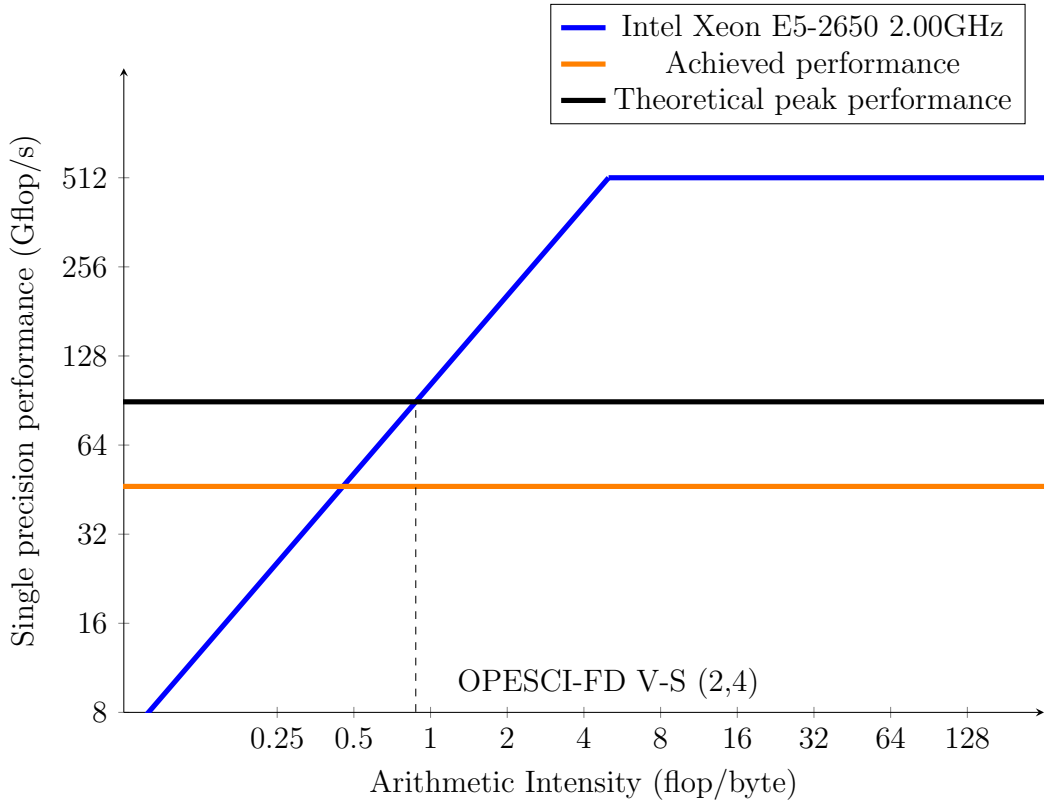

Figure 6.17: Roofline model of Intel Xeon E5-2650 and (2,4) velocity-stress FD kernel generated by OPESCI-FD

To calculate the total number of flops in the generated code, we use the number of flops in the kernel, multiply by the number of interior grid points and the number of time steps. In the example we used, there are 201 grid points per dimension (not including ghost cells), there are 192 flops in total in the velocity and stress kernels, and we run the simulation for 1000 time steps. This gives the total number of flops as $1.559 \times 10^{12}$. We compile the program with Intel® C++ compiler version 15.0, with optimisation level 3. We then time the program on CX1 using Intel® VTune™ running on 16 threads. The threads are pinned to the cores using `OMP_PROC_BIND=close`. This ensures the operating system completely fills one socket with the first 8 threads beforing using the second socket, thus minimising NUMA traffic. We obtain the achieved performance of 46.4 Gflops per second. This is 52% of the theoretical peak performance at this AI level.

We recognize this manual process is not ideal, as the ghost cell updates are ignored (which accounts for about 2% of the time spent in our experience), and also the compiler can apply optimisation to reduce the number of flops in the kernel. One current development of OPESCI-FD is to incorporate PAPI[9] to count the total number of flops produced by the program automatically. This will also confirm the AI of the kernel.

The complete roofline model is shown in Figure 6.17. The vertical dashed line indicates the velocity-stress (2,4) kernel generated by OPESCI-FD. This kernel is clearly in the memory-bandwidth bounded section on the chart.

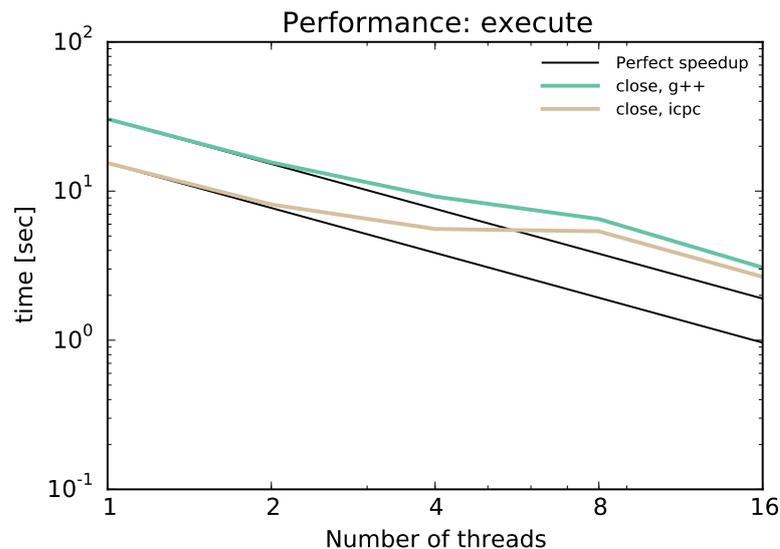

Figure 6.18: Parallel performance scaling on single node (Intel Xeon E5-2650)

Figure 6.18 shows how the program scales with increasing number of threads on a single CX1 Sandy Bridge node. We run a smaller simulation here, with only 100 grid points in each dimension. The program was compiled with Intel® C++ compiler version 15.0[10] and

---
[9]http://icl.cs.utk.edu/papi/
[10]Bundled with Intel® Parallel Studio XE 2015

GCC C++ compiler version 4.9.1, both set to level 3 optimisation. Again the threads are pinned to the cores using `OMP_PROC_BIND=close`. Hyper-threading (by using logical cores) is not applied. The results indicate good scalability until the program exhausts the memory bandwidth on a single socket, and the performance improves again when the program is running on two sockets.

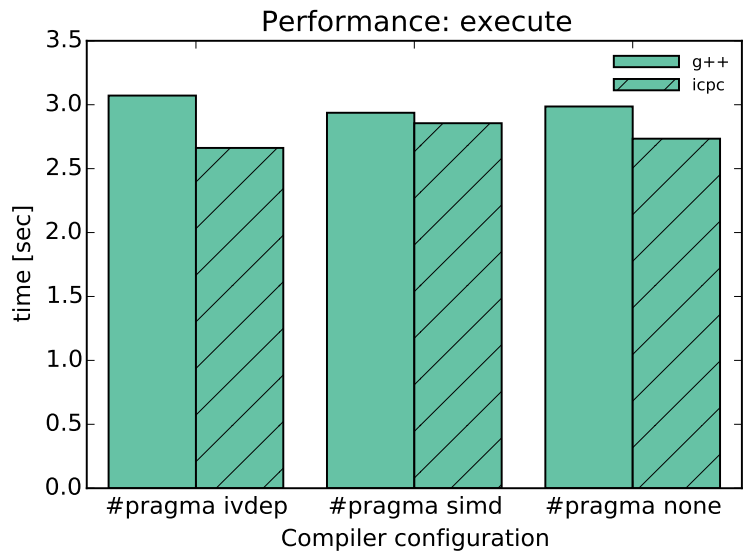

Figure 6.19: Performance comparison of preprocessor directives

One advantage of OPESCI-FD is the flexibility to testing the effects of different settings easily. We generate three similar C++ source code from the same StaggeredGrid object: the first with `ivdep=True`, *i.e.* `#pragma ivdep` is inserted before the inner loops; the second with `simd=True`, *i.e.* `#pragma simd` is inserted before the inner loop; the last one with both flags set to `False`. The result is shown in Figure 6.19, which indicates using the `simd` flag could help GCC compiler but might adversely affect Intel compiler. Such results might not be what the application developers expect and thus require further investigation, and using OPESCI-FD this can be tested easily without extra efforts.

We should note that the exact performance analysis and reasoning, while intellectually interesting, is not a focus at this stage of the project. OPESCI-FD aims to provide the infrastructure for automatic benchmarking on generated code, from which application devel-opers can draw their own conclusions. Functionality such as algorithmic AI calculation and the testing scripts in the current version of OPESCI-FD are built in for that purpose. Key features being added to the development version of OPESCI-FD currently include internalisation of compilation and automatic flops counting using PAPI. These features will further improve the efficiency and accuracy of benchmarking.

This page intentionally left blank.

# Chapter 7

# Conclusion and future work

In this chapter, we summarise the key features and advantages of OPESCI-FD and the lessons that we learnt when building OPESCI-FD. We conclude the report by discussing highlights of current development of OPESCI-FD and possible directions of future improvements.

## 7.1 Evaluation

We developed OPESCI-FD with the goal to alleviate the users from the burden of manual implementation of FD numerical methods. When applied to the velocity-stress FD schemes, which are commonly used in computational seismic studies, OPESCI-FD has demonstrated the following advantages:

- **Usability**

  The abstraction provided by OPESCI-FD allows application domain developers to focus on the mathematical aspects of the problems, with the details of implementation hidden away. The user still needs to write some code in Python to create the program, but such code is typical much shorter, more intuitive, and easier to write compared to manual implementation of the scheme. OPESCI-FD also provides scripts and pre-defined functions to facilitate development and testing.

- **Accuracy and robustness**

The internal plumbing of OPESCI-FD is built on symbolic mathematics representations, which is relatively easy to reason about and manipulate. The SymPy library provides tools to perform tasks such as solving linear equations and inverting matrices. These functions simplify the derivation of the computation kernel considerably. The code generation process is automatic, systematic and scalable. Complicated kernels can be generated without introducing extra complexity. In practice, developers are acutely aware of the trade-off between higher order accuracy schemes, which are faster because less grid points are required (and larger time steps can be used), and lower order accuracy schemes which are easier to implement. OPESCI-FD levels the ground of implementation challenges to allow the user to choose FD schemes purely based on their needs.

Kernels of high order FD schemes are typically lengthy and subject to subtle errors introduced by programmer, such as mis-spelling of variables and off-by-one errors in the array indices. These bugs can be hard to detect and tracked down. As discussed in Section 6.5, the eigenwave test case and visualisation functionality provided by OPESCI-FD are effective in identifying such errors. These techniques are general and can be applied to new FD schemes seamlessly. In our experience, the debugging process of the generated code is intuitive, and once an error has been identified, changes can be applied thoroughly to all places, which avoids creating new bugs when correcting old ones.

- **Flexibility**

The structures of FD schemes allow OPESCI-FD to create code segments which are assembled together later, using template files. Each fragment is created separately, so that various combinations of different parts can be experimented. For example, user can switch between different order of accuracy and 2D/3D grids simply by setting corresponding properties of the grid object. StaggeredGrid class also provides parameters to allow control of certain properties of the code, such as inserting vectorization directives. The tool chain is highly extensible. Application developers only need to extend the relevant classes to implement new schemes or generate code for new programming languages and stencil tools.

- **Possible performance improvement**

    OPESCI-FD provides facilities to benchmark performance of the generated code. By incorporating the code generation process and performance analysis in the same framework, OPESCI-FD encourages systematic exploration of new settings, which possibly lead to better performance on specific architectures.

- **Python community**

    OPESCI-FD is developed as an open-source Python package, which is easy to install and integrate with other Python libraries. The framework benefits from the powerful language features provided by the Python programming language, and also other open-source Python libraries, namely SymPy and Mako. Compared with stand-alone systems, the Python environment brings extra functionalities naturally, such as LaTeX printing of symbolic objects. Extra efforts are made to ensure the syntax and abstractions are intuitive to ease adoption, aided with documentations and examples. We hope that this will encourage more contributions from the wider computational numerical analysis community.

## 7.2 Limitations and future work

OPESCI-FD is very much a work in progress, and the flexibility of the tool chain opens doors for future extensions in many directions.

- **Implementation**

    The designs of OPESCI-FD is an iterative fine-tuning process which is constantly evolving. The current implementation still contains various rough edges and shortcomings. Currently the build process, implemented with CMake, is an independent component of the framework. A separate development is looking to internalise the build process

into the grid objects, so that compilation and testing can be achieved on the Python interface level.

- **Formatting of generated code**

    The code generated by OPESCI-FD lacks indentations and comments, resulting in low readability. Arguably the generated code is not intended for human readers, and this trade-off allows certain optimisation to happen, such as pre-valuing constants in the kernels. Nonetheless, in certain situations good readability is still much desired (*e.g.* when investigating the C++ compiler behaviour). However, it is not trivial to create correct indentations especially because OPESCI-FD generates all nested loops dynamically.

- **Additional FD schemes**

    The current version of OPESCI-FD only implements one FD scheme, namely the velocity-stress scheme on staggered grid. Despite the scheme's long-lasting popularity, it is nonetheless informative to explore other approaches in the context of seismic modeling, such as schemes based on the acoustic wave equations on a regular (non-staggered) grid. This enables users to take advantage of the flexibility of OPESCI-FD to compare performance and accuracy of different methodologies. It is natural to further generalise the grid representation for this purpose, such as creating a new Grid class which is parent class of all FD grids such as StaggeredGrid. However, the code generation library is unlikely to require significant amendments.

- **Auto-tuning**

    Deciding between different FD schemes for each application is a non-trivial task because of the trade-off between efficiency and accuracy of different schemes. Furthermore, each FD stencil can also be adjusted such that it is optimal for a certain frequency range to achieve higher accuracy [28]. With more schemes and settings added, iterative testing with the hope of deciding on a good combination of settings can become tedious. Some researchers have applied *Genetic Algorithms* [3] to explore the parameter space and

zoom in towards the likely optimal choice. Information from the roofline model also need to be incorporated in the selection process. Other than performance (in terms of Gflops/s), the optimisation goal can be other measures such as accuracy and energy consumption.

- **Downstream stencil and polyhedral compilers**

    So far we have only investigated code generation for Pochoir compiler (if not counting the "out-of-the-box" support of LLVM-Polly). Many stencil tools are actively being developed for HPC as discussed in Section 2.3. In particular, it is promising to investigate the tiling algorithms implemented by compilers using polyhedral models, such as Pluto [8]. For that purpose, the code generation library of OPESCI-FD needs to be extended.

- **Full-wave inversion**

    Full-wave inversion (FWI) is a technique to quantify subsurface physical properties from the data collected. FWI is the inverse problem of the propagator implemented by OPESCI-FD so far, and is extensively used in the oil and gas industry. We speculate that symbolic mathematics could play a role to simplify FWI code generation as well. Due to the non-linearity between observed data and model properties, FWI typically involves minimising an so-called objective function through iterative searching. Therefore real world FWI is both a Big Data problem and a HPC problem. The Python framework of OPESCI-FD would be highly suitable for integration into existing modern Big Data frameworks, such as Apache Spark[1], which is useful for FWI.

---

[1] http://spark.apache.org/